\documentclass[11pt]{article}

\usepackage[margin=1in]{geometry}
\usepackage{amsmath,amssymb,bm}
\usepackage{siunitx}
\usepackage{graphicx}
\usepackage{booktabs}
\usepackage{tabularx}
\usepackage{xcolor}
\usepackage{subcaption}
\usepackage{float}
\usepackage{placeins}
\usepackage[colorlinks=true,linkcolor=blue,citecolor=blue,urlcolor=blue]{hyperref}

\DeclareSIUnit{\torr}{torr}

\title{\textbf{Band-Selective LDOS Engineering of Yb/Er Upconversion:\\
       an Electromagnetic--Kinetic Diagnostic Framework}}

\author{%
\begin{minipage}{0.92\textwidth}
\centering
Yuxiang Zhang$^{1}$,\,
Mayte G\'omez-Casta\~no$^{2,3}$,\,
Agust\'in Mihi$^{3}$,\,
Serge Ravaine$^{2}$,\,
Xiaogang Liu$^{1}$,\,
Renaud A.~L.~Vall\'ee$^{2,\ast}$
\\[1.2ex]
\small
$^{1}$Department of Chemistry, National University of Singapore, 117543 Singapore\\
$^{2}$Univ.\ Bordeaux, CNRS, CRPP, UMR~5031, F-33600 Pessac, France\\
$^{3}$Institut de Ci\`encia de Materials de Barcelona (ICMAB-CSIC), Campus de la UAB, 08193 Bellaterra, Spain
\\[1ex]
$^{\ast}$Corresponding author: \href{mailto:renaud.vallee@u-bordeaux.fr}{renaud.vallee@u-bordeaux.fr}
\end{minipage}%
}
\date{}

\begin{document}
\maketitle

\begin{abstract}
 A persistent challenge in plasmonic upconversion is decoupling pump-field
 enhancement from emission-side local-density-of-optical-states (LDOS) 
 engineering to achieve selective band manipulation. Here, we show that
 a corrugated $\mathrm{SU_8}$/Au/$\mathrm{Al_2O_3}$ grating coated with
 a $\mathrm{NaYF_4:Yb/Er}$ upconversion nanoparticle (UCNP) monolayer
 realizes a truly band-selective platform. A broad plasmonic resonance
 near $670\,\mathrm{nm}$ modulates the red $\mathrm{Er}^{3+}$ decay
 rate by $\pm 15\%$ as a function of the $\mathrm{Al_2O_3}$ spacer
 thickness, while leaving the green transition experimentally invariant 
 ($<1\%$ change). Simultaneously, the $980\,\mathrm{nm}$ pump field 
 is monotonically suppressed below free-space levels, ensuring that 
 steady-state and time-resolved observables cleanly probe the 
 emission-side LDOS without pump interference.
 
 We analyze this system using a coupled electromagnetic--kinetic 
framework that integrates finite-difference time-domain (FDTD) calculations of Purcell factors and
pump fields with a six-level Yb/Er rate-equation model. The framework
quantitatively reproduces the $670\,\mathrm{nm}$ plasmonic resonance,
the red-band decay-rate modulation, and the monotonic decrease of the
green/red intensity ratio. Crucially, the model serves as a powerful
diagnostic tool: it over-predicts a green-band rate reduction, but
systematic parametric testing rules out geometric (apex smoothing) 
and material (grain-boundary damping, interband loss) imperfections 
as the cause. Instead, it isolates the residual discrepancy to 
measurement-versus-model factors (finite-aperture angular averaging)
 and missing non-radiative kinetic channels, establishing a clear 
 roadmap for the rational design and validation of future plasmonic-UCNP architectures.
\end{abstract}


\let\svaddcontentsline\addcontentsline
\renewcommand{\addcontentsline}[3]{}

\section{Introduction}
\label{sec:intro}

Lanthanide-doped upconversion nanoparticles (UCNPs), in particular
NaYF$_4$:Yb,Er systems, convert near-infrared (NIR) excitation into
visible anti-Stokes emission through sequential energy-transfer
upconversion (ETU) among long-lived $4f$ states.\cite{Fischer2011JAP,TorresVera2021,Wang2010Nature,Zhou2015NatNano}
Their distinctive photophysics underpins applications in
bioimaging, sensing, photovoltaics and anti-counterfeiting,\cite{Wu2019,Xu2021,Gnach2015,Goldschmidt2015,Xie2022}
yet the upconversion quantum yield (UCQY) under practical
irradiances remains modest, limited by weak absorption
cross-sections and the intrinsically slow radiative rates of
parity-forbidden $4f$--$4f$ transitions.

Nanophotonic engineering offers two complementary routes to
overcome these limitations.
First, near-field enhancement at the pump wavelength can
multiply the effective Yb$^{3+}$ absorption rate.
Second, tailoring the LDOS at the emission wavelengths accelerates
radiative decay and, through the antenna efficiency
$\eta_a = P_{\mathrm{rad}}/P_{\mathrm{tot}}$, can improve out-coupling
while limiting metal-induced quenching.\cite{Anger2006,NovotnyHecht}
Both mechanisms act simultaneously and interdependently, and a
recurring obstacle to rational design is that the same plasmonic
feature that boosts the pump field also tends to spectrally
overlap -- and thereby couple to -- one of the emission bands,
making the two enhancement channels difficult to address
independently.

Significant progress has been made on each piece of this problem.
Esteban \textit{et al.} established that upconversion is an
incoherent, population-governed process requiring a rate-equation
treatment, and identified pump-field enhancement and LDOS-modified
decay rates as the two key electromagnetic inputs to such a
model.\cite{Esteban2008}
The competition between near-field excitation enhancement and
metal-induced nonradiative quenching as a function of
emitter--metal separation was demonstrated for single molecules
by Anger \textit{et al.},\cite{Anger2006} and underlies the need
for a dielectric spacer in any ensemble platform.
Fischer \textit{et al.} provided accurate Einstein coefficients
for $\beta$-NaYF$_4$:Er$^{3+}$,\cite{Fischer2011JAP} and the
Goldschmidt group coupled these to LDOS calculations in
one-dimensional photonic crystals, achieving quantitative agreement
with macroscopic upconversion measurements.\cite{Herter2013,Hofmann2016,Hofmann2021}
On the plasmonic side, Fischer \textit{et al.} modelled
Purcell enhancement near gold nanospheres,\cite{Fischer2011OE,Fischer2016,Schietinger2010}
while Liu and Lei exploited double-resonant gold nanorods to
enhance both pump and emission simultaneously,\cite{LiuLei2015}
illustrating that the standard route to combined enhancement
relies on dual-resonance antennas.
More recently, we demonstrated giant upconversion
enhancement in deep-subwavelength Au nanotrenches by combining
FDTD Purcell maps with a kinetic model,\cite{Xu2021,Xu2024Chem}
and we achieved a 166-fold spontaneous-emission-rate
increase and emission superburst in a gap-mode plasmonic
nanocavity.\cite{Wu2019,Chen2022NatPhoton}
Torres Vera \textit{et al.} provided an advanced kinetic model
for the Yb/Er dopant-ratio and pump-intensity dependence of the
green/red emission balance.\cite{TorresVera2021}

These advances notwithstanding, two questions remain open at
the ensemble scale relevant for applications.
\textit{(i)}~Can a fabricable platform deliver a clean
\emph{band-selective} LDOS modification -- one that engineers
the decay of one Er$^{3+}$ transition while leaving the other
untouched, without pump-side coupling that would obscure the
spectroscopic signature of the LDOS?
\textit{(ii)}~Can a coupled electromagnetic--kinetic framework
predict the multiple observables of an ensemble UCNP platform
(steady-state spectra, green/red ratio, time-resolved decays)
quantitatively, and -- when it fails -- identify which
photophysical or geometric ingredient must be revised?
A unified framework addressing both questions would establish
design rules connecting a fabrication parameter to spectroscopic
observables, and bridge the gap between extreme single-emitter
Purcell regimes\cite{Wu2019,Chen2022NatPhoton} and macroscopic
ensemble platforms.

In this work we address both questions on a corrugated
SU8/Au/Al$_2$O$_3$ grating coated with a dense
NaYF$_4$:Yb(20\%),Er(5\%) UCNP monolayer.
A broad plasmonic extinction resonance centred near
\SI{670}{\nano\meter} -- spectrally aligned with the red
$^4F_{9/2} \to {}^4I_{15/2}$ Er$^{3+}$ transition -- modifies
the radiative and total LDOS at the red emission wavelength as
the Al$_2$O$_3$ spacer thickness $d$ is varied between 5 and
\SI{25}{\nano\meter}, producing decay-rate variations of up to
$\pm$15\% with respect to the quartz reference.
The green $^2H_{11/2}/^4S_{3/2} \to {}^4I_{15/2}$ transition,
spectrally separated from the resonance, is essentially
unperturbed: experimental green-band lifetimes are invariant
within $\pm 1\%$ across all $d$, providing a clean
band-selective design lever.
Crucially, the pump field at \SI{980}{\nano\meter} is
\emph{not} resonantly enhanced on this platform: averaged over
the UCNP positions sampled along the central bump,
$\langle f_{\rm exc}\rangle$ remains below unity for all
spacer thicknesses ($\langle f_{\rm exc}\rangle = 0.27$--$0.48$
between $d = 5$ and \SI{25}{\nano\meter}), so spectroscopic
observables in the steady state and in time-resolved
photoluminescence cleanly probe the LDOS-side resonance without
pump-side interference.

Having established this band-selective lever experimentally, our next
aim is to account for it quantitatively: to build a model transparent
enough to say which of the observed effects the photonic engineering
explains, and which it does not. To this end, we accompany the
experimental observations with a coupled electromagnetic--kinetic
framework that has not, to our knowledge, been demonstrated at this
level of integration for ensemble UCNP platforms.

The framework combines orientation- and surface-averaged
radiative and total Purcell factors at \SI{550}{\nano\meter}
and \SI{660}{\nano\meter}, the local pump-field enhancement
$f_{\mathrm{exc}}$ at \SI{980}{\nano\meter}, and a simplified
six-level Yb/Er rate-equation model that explicitly separates
radiative, intrinsic nonradiative and environment-induced
nonradiative decay channels.
The framework reproduces the linear extinction resonance at
\SI{670}{\nano\meter}, the amplitude of the red-band
decay-rate modulation ($\pm 10$--$15\%$), and the monotonic
decrease of the simulated green/red intensity ratio with $d$.
It does \emph{not} reproduce two features of the data:
the absolute magnitude of the
green/red ratio (over-predicted by a factor $\sim 1.5$); and
the experimentally invariant green-band lifetime (the
simulation predicts a $\sim$25--30\% reduction).
The simulated red-band decay-rate enhancement is monotonic
in $d$ and agrees with the measured experimental data
(uniform $\approx +15\%$ across all $d$; see
Sec.~\ref{sec:trpl}).
We test the accessible corrections to the green-band
over-prediction in turn -- ridge-tip smoothing
($h_{\mathrm{round}} \in \{0, 5, 10\}~\si{\nano\meter}$),
grain-boundary (free-electron) damping of the gold,
geometric ridge disorder, and enhanced interband loss
(SI Sec.~\hyperref[sec:SI_extended_green]{S4.2}) -- and find
that none closes the residual: smoothing and disorder change
the surface-averaged radiative Purcell factor by only a few
percent, while grain-boundary damping acts predominantly on
the red and interband loss moves the green feature the wrong
way.
The framework therefore serves a diagnostic role: it captures
what depends on the broad, geometrically robust red-band
plasmonic resonance, and isolates the green-band over-prediction
as a residual that survives every accessible electromagnetic
and material correction, pointing instead to
measurement-versus-model factors (finite-aperture angular
averaging) and to additional non-radiative channels at the
green transition that the present six-level kinetic model
does not include.

The paper is organised as follows.
Section~\ref{sec:platform} presents the nanophotonic platform
and its linear optical response.
Section~\ref{sec:ldos} reports the LDOS modification at the
green and red Er$^{3+}$ transitions and the monotonic
suppression of the pump field at \SI{980}{\nano\meter}.
Section~\ref{sec:steady_state} compares experimental and
simulated steady-state spectra and green/red ratios and
presents the power-dependence of the upconversion enhancement.
Section~\ref{sec:trpl} provides the time-resolved validation,
including the multi-observable test of the framework, the
ridge-tip-smoothing diagnostic, and an explicit discussion
of the green-band discrepancy.
Section~\ref{sec:conclusions} extracts design rules and
outlooks the framework's extension to other architectures.

\section{Results and Discussion}
\label{sec:results}

\subsection{Nanophotonic Platform and Linear Optical Response}
\label{sec:platform}

We seek a structure that targets one goal: place a strong, broad
plasmonic resonance right on the red Er$^{3+}$ emission band
($\sim$\SI{660}{\nano\meter}), while keeping the green band and the
\SI{980}{\nano\meter} pump well off-resonance. This way we can boost
the red channel selectively, leaving the green untouched as an
internal reference. The geometry below follows directly from this aim.

\begin{figure}[H]
  \centering
  \includegraphics[width=\textwidth]{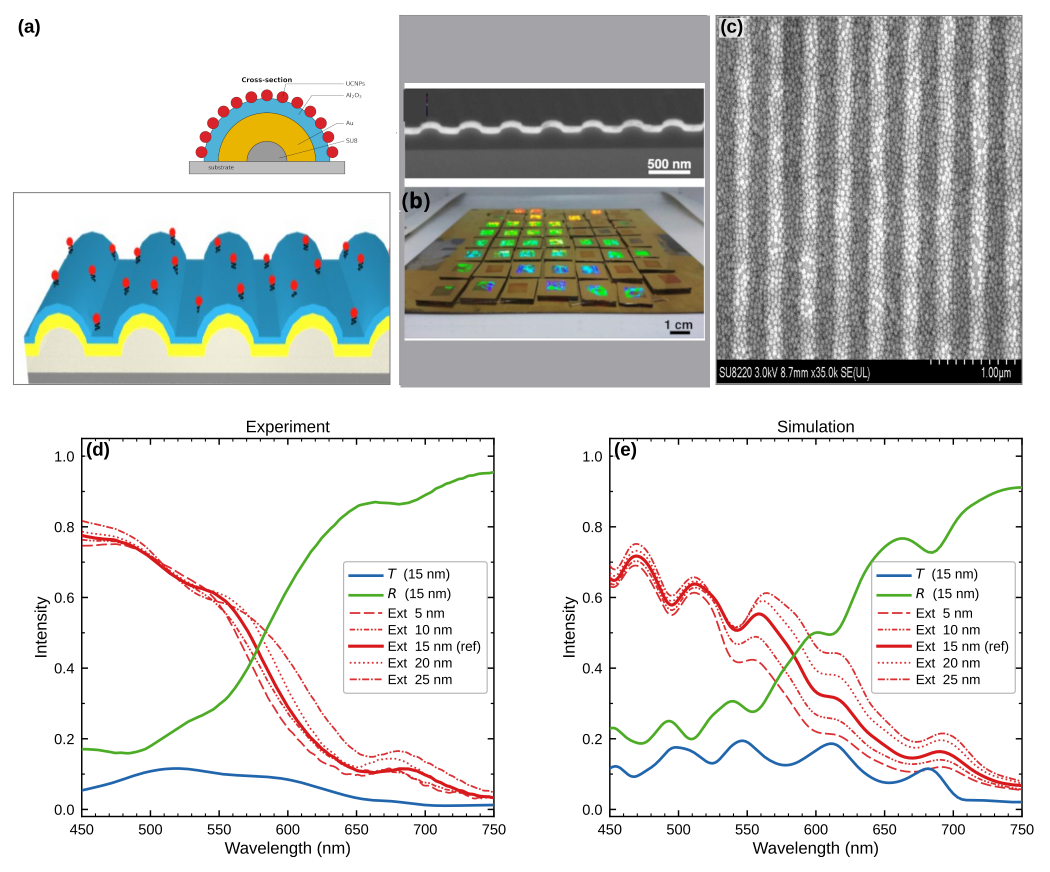}
  \caption{%
    \textbf{Nanophotonic platform.}
    (a)~Schematic of the corrugated SU8/Au/Al$_2$O$_3$ grating
    with spin-coated UCNP monolayer.
    (b)~Cross-sectional SEM image of the SU8/Au/Al$_2$O$_3$ stack
    (scale bar \SI{500}{\nano\meter}) and photograph of the
    sample array (scale bar \SI{1}{\centi\meter}).
    (c)~Top-view SEM of the dense UCNP monolayer
    (scale bar \SI{1}{\micro\meter}).
    (d)~Measured and (e)~simulated transmission $T$, reflectance
    $R$ and extinction $\mathrm{Ext} = 1-T-R$ for the five spacer
    thicknesses.
    The structure is a broadband mirror with a plasmonic
    extinction maximum at $\lambda \approx \SI{670}{\nano\meter}$
    aligned with the red $^4F_{9/2} \to {}^4I_{15/2}$ transition;
    the resonance position is robust against $d$ while its
    amplitude grows with $d$. The weaker measured
    $\sim$\SI{550}{\nano\meter} feature is analysed in
    SI Sec.~\hyperref[sec:SI_extended_green]{S4.2}.
    See Sec.~\ref{sec:platform}.
  }
  \label{fig:platform}
\end{figure}

\paragraph{Design rationale.}
The geometry follows from three decoupled design parameters
mapped onto three distinct targets. (i)~The \emph{grating period}
$\Lambda$ sets the spectral position of the plasmonic resonance:
to first order $\lambda_{\rm res} \approx \Lambda\, n_{\rm eff}$
with $n_{\rm eff} \approx 1.6$--$1.7$ the effective index of the
metal/dielectric mode, so $\Lambda = \SI{400}{\nano\meter}$ places
the broad resonance at $\approx 660$--\SI{680}{\nano\meter},
deliberately aligned with the red
$^4F_{9/2} \to {}^4I_{15/2}$ transition and separated from both
the green band and the \SI{980}{\nano\meter} pump. Band
selectivity is thus a designed consequence: placing the
resonance on the red band leaves the green transition
off-resonance (and near the gold interband edge, where the
plasmonic response is intrinsically weak). (ii)~The \emph{ridge
amplitude} $b_{\rm SU8} \approx \SI{35}{\nano\meter} \ll \Lambda$
supports a well-defined resonance with field overlap at the
emitter layer while avoiding spurious sub-wavelength hot-spots.
(iii)~The \emph{spacer thickness} $d$ is the tuning knob: it does
not shift the resonance ($\Delta\lambda < \SI{5}{\nano\meter}$
across $d = 5$--\SI{25}{\nano\meter}, Fig.~\ref{fig:platform}d,e)
but controls the spatial overlap between the plasmonic mode and
the UCNP monolayer, hence the radiative-versus-ohmic balance.

The nanophotonic platform is fabricated by UV laser interference
lithography of an SU8 photoresist into parallel ridges with period
$a = \SI{400 \pm 5}{\nano\meter}$ and amplitude
$\sim\SI{35 \pm 3}{\nano\meter}$ on a fused-silica substrate,
followed by thermal evaporation of a \SI{50}{\nano\meter} Au film
and atomic layer deposition (ALD) of a conformal Al$_2$O$_3$
spacer of thickness $d$ between 5 and \SI{25}{\nano\meter}
(Fig.~\ref{fig:platform}a).
The cross-sectional scanning electron microscopy (SEM) image of the
SU8/Au/Al$_2$O$_3$ stack is shown in Fig.~\ref{fig:platform}b.
NaYF$_4$:Yb(20\%),Er(5\%) UCNPs (hexagonal $\beta$ phase,
lateral dimension $\sim$\SI{22 \pm 3}{\nano\meter}, height
$\sim$\SI{15 \pm 2}{\nano\meter}) are deposited by spin-coating
to form a dense, nearly close-packed monolayer that conforms to
the grating topography (Fig.~\ref{fig:platform}c).
Full fabrication details are reported in
SI Sec.~\hyperref[sec:SI_fab]{S1}.

The measured linear optical response (Fig.~\ref{fig:platform}d)
confirms that the structure acts as an opaque broadband mirror
($T < 0.15$, $R > 0.6$ above \SI{600}{\nano\meter}) with a narrow
extinction maximum at $\lambda \approx \SI{670}{\nano\meter}$
that spectrally coincides with the red
$^4F_{9/2} \to {}^4I_{15/2}$ Er$^{3+}$ transition.
The position of the resonance is essentially independent of $d$
across the full \SI{5}{\nano\meter}--\SI{25}{\nano\meter} range,
with only a $\sim$\SI{5}{\nano\meter} shift between the thinnest
and thickest spacers.
FDTD simulations reproduce both the resonance position and the
\textit{trend} of its amplitude vs.\ $d$ (Fig.~\ref{fig:platform}e),
confirming that the \SI{670}{\nano\meter} feature is a genuine
plasmonic resonance of the corrugated metal--dielectric stack
rather than an interference artefact.
A secondary extinction feature predicted near \SI{550}{\nano\meter}
is markedly weaker in the experimental spectra, with a discrepancy
of roughly a factor of two in amplitude.
This green-side mismatch is examined in detail in
SI Sec.~\hyperref[sec:SI_extended_green]{S4.2}: at first glance, it should trace back
to a combination of fabrication- and material-related effects
that suppress short-wavelength near-field features without
significantly affecting the longer-wavelength resonance at
\SI{670}{\nano\meter}.
The candidate corrections include smoothing of the ridge tips
(the experimental profile is closer to a low-amplitude sinusoid
than to the idealised half-ellipse profile used in the FDTD
model), grain-boundary damping in the evaporated gold,
enhanced interband loss, and
inhomogeneous broadening from ridge-position jitter; we test
each of these explicitly in
Sec.~\ref{sec:trpl} and SI~\hyperref[sec:SI_extended_green]{S4.2}
and find that none accounts for the green-side mismatch.
The quantitative reproduction of the red-band resonance
nevertheless gives confidence that the simulated LDOS at
\SI{660}{\nano\meter} is reliable, whereas predictions at
\SI{550}{\nano\meter} should be treated as semi-quantitative --
a point that we revisit and exploit explicitly in
Sec.~\ref{sec:trpl} as a diagnostic feature of the framework.

We emphasise that grating-based plasmonic structures have been
used previously to enhance UCNP emission through surface plasmon polariton (SPP) coupling at
the pump wavelength;\cite{Le2024} our platform instead engineers
the LDOS at the emission wavelength, with the corrugation acting
as a broadband mirror at \SI{980}{\nano\meter} that suppresses
the pump field rather than enhancing it (Sec.~\ref{sec:ldos}).

\subsection{Band-Selective LDOS at the Red Er$^{3+}$ Transition
            and Pump-Field Suppression}
\label{sec:ldos}

Before modelling the emission, we first verify that the fabricated
structure does what it was designed to do: enhance the photonic
environment selectively on the red band, leave the green band weakly
affected, and boost the pump field at \SI{980}{\nano\meter}. We probe
this directly through the local density of optical states (LDOS) at
the two emission wavelengths and the pump-field intensity at the
excitation wavelength.

\begin{figure}[H]
  \centering
  \includegraphics[width=\textwidth]{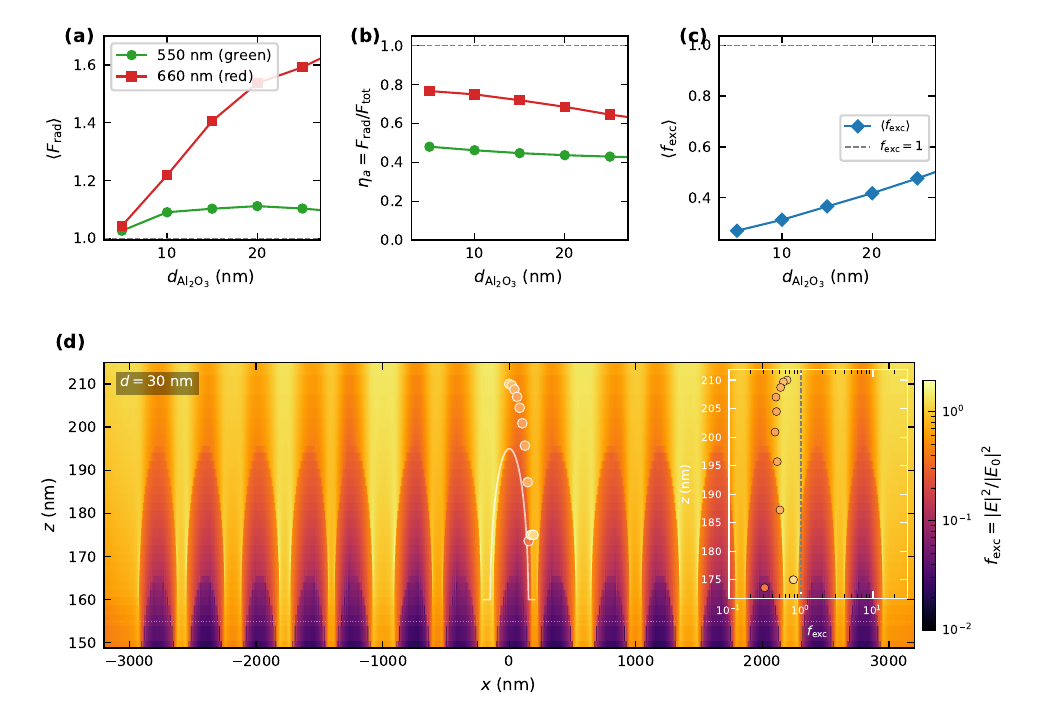}
  \caption{%
    \textbf{LDOS modification and pump-field suppression at the
    UCNP layer.}
    (a)~Surface-averaged radiative Purcell factor
    $\langle F_{\mathrm{rad}}\rangle$ at \SI{550}{\nano\meter}
    (green circles) and \SI{660}{\nano\meter} (red squares) vs.\
    $d$: the red band grows monotonically with the
    \SI{670}{\nano\meter} resonance
    ($\langle F_{\mathrm{rad}}\rangle \approx 1.0 \to 1.6$) while
    the green stays near unity (1.02--1.11), confirming spectral
    isolation.
    (b)~Antenna efficiency
    $\eta_a = F_{\mathrm{rad}}/F_{\mathrm{tot}}$, lower at the
    green band ($\approx 0.43$--$0.48$) than at the red
    ($\approx 0.61$--$0.77$), reflecting stronger metal-induced
    quenching off resonance.
    (c)~Mean pump-field enhancement
    $\langle f_{\mathrm{exc}}\rangle$ at the UCNP positions:
    suppressed below the free-space reference at all $d$
    ($\approx 0.27 \to 0.50$ from $d = 5$ to \SI{30}{\nano\meter}),
    with no resonant enhancement.
    (d)~Pump-field map $f_{\mathrm{exc}}(x,z)$ at
    \SI{980}{\nano\meter} for $d = \SI{30}{\nano\meter}$
    (logarithmic scale): laterally homogeneous and sub-unity at
    the emitter positions. Coloured circles: UCNP sampling
    positions; inset: $z$ vs.\ $f_{\mathrm{exc}}$.
    Maps at all $d$ in SI
    Sec.~\hyperref[sec:SI_extended_maps]{S4.5}.
    See Sec.~\ref{sec:ldos}.
  }
  \label{fig:ldos_purcell}
\end{figure}

To quantify the electromagnetic environment experienced by the
UCNP monolayer, we computed surface- and orientation-averaged
radiative Purcell factors $F_{\mathrm{rad}}$ and total Purcell
factors $F_{\mathrm{tot}}$ at \SI{550}{\nano\meter} (green band)
and \SI{660}{\nano\meter} (red band), the corresponding antenna
efficiency $\eta_a = F_{\mathrm{rad}}/F_{\mathrm{tot}}$, and the
local pump-field enhancement
$f_{\mathrm{exc}}(\mathbf{r}) = |E(\mathbf{r})|^2/|E_0(\mathbf{r})|^2$
at \SI{980}{\nano\meter}, all relative to a flat SU8 substrate
reference.
Sampling positions follow the Al$_2$O$_3$ outer surface profile
of the central bump, with each dipole placed at a distance
$r_{\mathrm{UCNP}} = \SI{15}{\nano\meter}$ above the surface
along the local outward normal (11 positions per simulation,
see SI Sec.~\hyperref[sec:SI_FDTD]{S3.2}).
The LDOS quantities are evaluated on the original 5-point
spacer sweep ($d = 5, 10, 15, 20, \SI{25}{\nano\meter}$); the
pump-field calculation is extended by one point at
$d = \SI{30}{\nano\meter}$ to confirm the asymptotic behaviour
of $\langle f_{\mathrm{exc}}\rangle$ at large $d$.
The averaged results are summarised in
Fig.~\ref{fig:ldos_purcell}.

\paragraph{Emission-side resonance and band selectivity.}
The red-band $\langle F_{\mathrm{rad}}\rangle$
(Fig.~\ref{fig:ldos_purcell}a, red squares) increases monotonically
with $d$ from $\langle F_{\mathrm{rad}}\rangle \approx 1.0$ at
$d = \SI{5}{\nano\meter}$ to $\approx 1.6$ at
$d = \SI{25}{\nano\meter}$.
This trend reflects the growing modal volume of the air-side
plasmon as the dielectric load increases: at small $d$ the UCNPs
are strongly coupled to the plasmonic mode but most of the
emitted power is dissipated as ohmic loss in the metal, yielding
low $F_{\mathrm{rad}}$ and the lowest antenna efficiency
($\eta_a^{660} \approx 0.77$ at $d = \SI{5}{\nano\meter}$); as
$d$ grows, the radiative weight of the mode increases more
rapidly than the ohmic-loss channel, raising $F_{\mathrm{rad}}$
while $\eta_a^{660}$ decreases moderately (from 0.77 to 0.61
between $d = 5$ and \SI{25}{\nano\meter}, Fig.~\ref{fig:ldos_purcell}b).
The decrease of $\eta_a$ with $d$ may seem counter-intuitive
compared to the $1/d^n$ recovery expected for a single emitter
above a planar mirror, but it reflects the specific modal
structure of the corrugated mirror in the spacer range probed
here, where all UCNPs remain within the near-field of the
plasmonic mode (see SI~Sec.~\hyperref[sec:SI_extended_convergence]{S4.1}
for a full discussion).

By contrast, $\langle F_{\mathrm{rad}}^{550}\rangle$ remains
within $1.02$--$1.11$ of unity for all $d$, with a shallow,
non-monotonic dependence on $d$ peaking near
$d = \SI{20}{\nano\meter}$ (Fig.~\ref{fig:ldos_purcell}a, green
circles).
The near-unity behaviour at \SI{550}{\nano\meter} confirms that
the \SI{670}{\nano\meter} plasmonic resonance does not spectrally
overlap the green band -- the central condition for the
band-selective LDOS engineering claimed in the abstract and
supported experimentally in Sec.~\ref{sec:trpl}.
The systematically lower $\eta_a^{550} \approx 0.43$--$0.48$ at
the green band reflects the off-resonant character of this
emission relative to the plasmonic feature, which preferentially
channels the green-band emitted power into ohmic loss rather
than propagating radiation.

\paragraph{Pump-side suppression at \SI{980}{\nano\meter}.}
Figure~\ref{fig:ldos_purcell}c reports
$\langle f_{\mathrm{exc}}\rangle$ at the UCNP positions as a
function of $d$.
On this platform, the corrugated Au mirror suppresses the local
pump intensity below the free-space reference at all spacer
thicknesses: $\langle f_{\mathrm{exc}}\rangle$ grows monotonically
from $\approx 0.27$ at $d = \SI{5}{\nano\meter}$ to $\approx 0.48$
at \SI{25}{\nano\meter}, and asymptotes to $\approx 0.50$ at
$d = \SI{30}{\nano\meter}$.
This monotonic suppression is consistent with the broadband
mirror response at \SI{980}{\nano\meter}: at the apex of the
bumps, the UCNP layer sits within the first
node--antinode region of the standing wave above the metal
surface, where the field is partially suppressed relative to a
flat dielectric substrate.
As $d$ increases the UCNPs drift further from the metal,
recovering progressively the free-space field; $\langle
f_{\mathrm{exc}}\rangle$ remains below unity over the entire
investigated range, however, indicating that
\SI{30}{\nano\meter} of Al$_2$O$_3$ is still insufficient for
the emitter layer to escape the suppression.
The pump-field map at $d = \SI{30}{\nano\meter}$
(Fig.~\ref{fig:ldos_purcell}d) confirms the spatially
homogeneous, sub-unity character of $f_{\mathrm{exc}}$ above
the bumps; the weak vertical striations periodic with the
grating reflect the $\sim$\SI{400}{\nano\meter}-period
modulation of the air-side standing wave, not gap-confined
plasmonic modes.
Maps at all spacer thicknesses (SI
Sec.~\hyperref[sec:SI_extended_maps]{S4.5}) confirm the absence
of localised hotspots at the UCNP positions.

\paragraph{Implication for the spectroscopic observables.}
The combined picture from Fig.~\ref{fig:ldos_purcell}a--c is
that the Al$_2$O$_3$ spacer controls a single emission-side
LDOS resonance, spectrally aligned with the red Er$^{3+}$ band
and decoupled from the pump channel.
This is the defining feature of the platform from a design
perspective: spectroscopic observables (steady-state spectra,
green/red ratio, time-resolved decays) probe directly the LDOS
at the red band, with no overlapping pump-side enhancement
that would otherwise have to be deconvolved.
The $\sim$3$\times$ variation of $\langle f_{\mathrm{exc}}\rangle$
across the investigated $d$ range does enter the steady-state
intensities through the Yb$^{3+}$ ground-state absorption rate,
but in the saturation regime accessed by the experiments
(Sec.~\ref{sec:steady_state}, $I_{\mathrm{pump}} \sim
1$--\SI{10}{\kilo\watt\per\centi\meter\squared}), this variation
contributes only a moderate, monotonic offset to the predicted
emission enhancements, leaving the LDOS-driven features of the time-resolved decays cleanly accessible.

\subsection{Steady-State Upconversion and Green/Red Ratio}
\label{sec:steady_state}

Having characterised the engineered photonic environment, we now ask
how it reshapes the actual upconversion emission. The steady-state
spectra reveal the net green/red intensity balance and provide the
ratio that the combined electromagnetic--kinetic model must reproduce.

\begin{figure}[H]
  \centering
  \includegraphics[width=\textwidth]{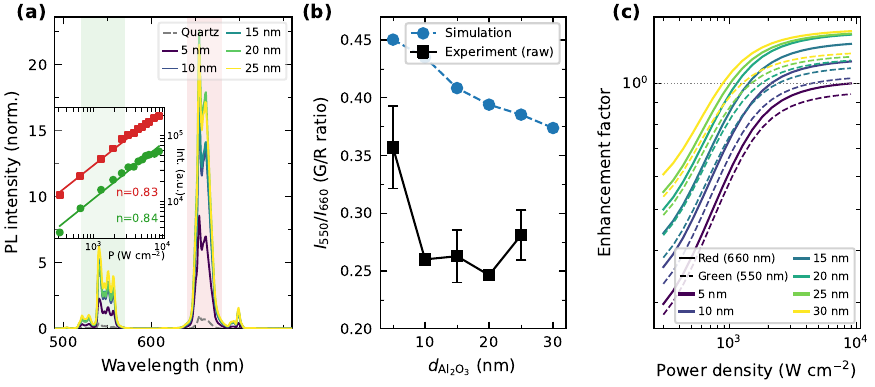}
  \caption{%
    \textbf{Steady-state upconversion emission.}
    (a)~Mean PL spectra ($d = 5$--\SI{25}{\nano\meter},
    \SI{980}{\nano\meter} CW,
    $I_{\rm pump} = \SI{9}{\kilo\watt\per\centi\meter\squared}$);
    dashed grey: quartz reference. Inset: integrated band
    intensities vs.\ pump density
    ($n_{\rm green} = 0.84$, $n_{\rm red} = 0.83$, indicating
    above-saturation operation).
    (b)~Green-over-red ratio $I_{550}/I_{660}$ vs.\ $d$:
    experiment (black squares, 
    $1\sigma$ over three replicates per thickness) and coupled
    EM--kinetic simulation (blue circles).
    Both peak at $d = \SI{5}{\nano\meter}$; the experiment is flat
    within scatter thereafter and the simulation over-predicts the
    absolute ratio by a factor $\sim$1.5.
    (c)~Simulated emission enhancement vs.\ pump-power density
    (solid: red band; dashed: green band); the enhancement
    decreases with irradiance as the structured and reference
    samples saturate at different effective rates.
    See Sec.~\ref{sec:steady_state}.
  }
  \label{fig:emission}
\end{figure}

Upconversion photoluminescence (PL) spectra were recorded under
\SI{980}{\nano\meter} continuous-wave (CW) excitation at power densities of
300--\SI{9000}{\watt\per\centi\meter\squared} for all five
spacer thicknesses.
The spectra (Fig.~\ref{fig:emission}a) display the structured
green band ($^2H_{11/2}/^4S_{3/2} \to {}^4I_{15/2}$,
520--\SI{570}{\nano\meter}) and the dominant red band
($^4F_{9/2} \to {}^4I_{15/2}$, 640--\SI{680}{\nano\meter}),
both of which grow with pump power.
Logarithmic slopes of the integrated band intensities vs.\
pump density (inset) are
$n_{\rm green} = 0.84$ and
$n_{\rm red} = 0.83$, well below the value $n = 2$ characteristic
of unsaturated two-photon ETU and consistent with operation
above the Yb$^{3+}$ saturation threshold
$I_{\rm sat} \approx \SI{3}{\kilo\watt\per\centi\meter\squared}$
estimated from the absorption cross-section
$\sigma_Y = \SI{1.7e-20}{\centi\meter\squared}$ and excited-state
lifetime $\tau_Y = \SI{2}{\milli\second}$
(SI Sec.~\hyperref[sec:SI_theory]{S2.3}).\cite{TorresVera2021,Park2015}

\paragraph{Choice of Er concentration.}
The 5\% Er doping is an intentional choice. A high Er content
raises the Yb$\to$Er energy-transfer coefficient
(approximately linearly in $[\mathrm{Er}]/[\mathrm{Yb}]$),
which is needed to operate at and above the Yb saturation
threshold over the experimental range and to obtain bright,
high-signal-to-noise decay traces -- essential, since the
observable of interest is a $\pm 15\%$ LDOS-induced change in
the red-band decay rate.
The accompanying Er--Er cross-relaxation preferentially
populates the red $^4F_{9/2}$ manifold, yielding the
red-dominant spectra ($I_{550}/I_{660} < 1$) that make the red
band -- on which the plasmonic resonance is engineered -- the
strongest, cleanest decay channel, while the off-resonant green
band serves as an internal control. Self-quenching at 5\% is
not neglected but absorbed into the effective intrinsic
non-radiative rates $\gamma^0_{\mathrm{nr},2}$ and
$\gamma^0_{\mathrm{nr},3}$, calibrated directly on the measured
lifetimes of the actual UCNPs on quartz
(SI Table~S1), so the model is parameterised on the real
concentration-dependent photophysics.
At a lower concentration (e.g.\ 2\%), reduced cross-relaxation
would lengthen both intrinsic lifetimes and raise the green/red
ratio,\cite{TorresVera2021} whereas the \emph{relative} LDOS
modulation $k/k_{\rm ref}$ -- set by the nanostructure's Purcell
factors and independent of the dopant kinetics -- would be
preserved, so the band-selective engineering carries over; the
main practical penalty would be a lower absolute red intensity
and reduced decay-curve signal-to-noise.

\paragraph{Green/red ratio: monotonic decrease in simulation,
flat plateau in experiment.}
The experimental green-over-red ratio $I_{550}/I_{660}$
(Fig.~\ref{fig:emission}b, black squares; processed from the
raw spectra with background
subtraction, fixed integration windows, three replicates
averaged per thickness --- see SI
Sec.~\hyperref[sec:SI_extended_decays]{S4.3}) is highest for the
thinnest spacer ($d = \SI{5}{\nano\meter}$, $I_{550}/I_{660}
\approx 0.36$), consistent with the direction of the simulated
trend, and is otherwise essentially flat within the
replicate-to-replicate scatter across $d = 10$--\SI{25}{\nano\meter}
($I_{550}/I_{660} \approx 0.25$--$0.28$).
The residual non-monotonicity, including the slightly lower
value near $d = \SI{20}{\nano\meter}$, lies within the
measurement scatter (error bars in Fig.~\ref{fig:emission}b)
and is not interpreted as a physical feature.
The coupled EM--kinetic model (blue circles) reproduces the
overall decreasing trend across the full $d$ range and gives a
ratio of the right order ($\sim 0.37$--$0.46$ predicted vs.\
$0.25$--$0.36$ measured), the $d = \SI{5}{\nano\meter}$ maximum
being common to both.
Successfully capturing the order of magnitude and dominant 
decreasing trend provides a non-trivial benchmark for the 
framework. The remaining quantitative gap—an absolute 
$\sim 1.3$--$1.5\times$ overestimation of the ratio--mirrors the 
over-predicted green-band features at $\sim\SI{550}{\nano\meter}$ in the simulated linear   
response (Fig.~\ref{fig:platform}e). This short-wavelength discrepancy 
traces back to the same geometric and material-related corrections 
discussed in Sec.~\ref{sec:platform}, analysed for the green band in 
Sec.~\ref{sec:trpl}, and detailed in SI Sec.~\hyperref[sec:SI_extended_green]{S4.2}.

\paragraph{Power-dependent enhancement.}
The simulated emission enhancement vs.\ pump-power density
(Fig.~\ref{fig:emission}c) exhibits the characteristic decrease
expected when the structured and reference samples saturate at
different effective rates: at low irradiances the structured
sample, with its higher effective absorption rate (modulated by
the LDOS at the emission wavelengths and by the reduced but
$d$-dependent pump field, Sec.~\ref{sec:ldos}), saturates first,
while the flat reference still operates in the unsaturated
regime; as the pump density increases, both samples approach
the same population-limited ceiling and the relative
enhancement decreases.
In the practically relevant range
$\sim 1000$--\SI{10000}{\watt\per\centi\meter\squared}, the
predicted enhancement crosses unity around
\SI{1}{\kilo\watt\per\centi\meter\squared} for
$d \ge \SI{15}{\nano\meter}$ and reaches 2--3$\times$ at the
high-irradiance end of the range, while it remains below unity
for $d = \SI{5}{\nano\meter}$, where metal-induced quenching is
largest.
The experimentally measured enhancement factors at the same
power densities are 10--35$\times$ -- larger than the simulated
values by roughly an order of magnitude.
We attribute this difference to the choice of experimental
reference (UCNPs on bare quartz, with a lower-index environment
and smaller LDOS baseline than the flat-SU8 reference used in
the simulation) and to the additional anti-reflection coating
effect of the corrugated stack at \SI{980}{\nano\meter}, neither
of which is captured by the flat-SU8 reference geometry.
The relative ordering of the $d$ values and the qualitative
power dependence are nevertheless faithfully reproduced.
A full set of power-dependence curves and the analysis of the
reference baseline are reported in
SI Sec.~\hyperref[sec:SI_extended_power]{S4.4}.

\subsection{Time-Resolved Decay Dynamics}
\label{sec:trpl}

\begin{figure}[H]
  \centering
  \includegraphics[width=\textwidth]{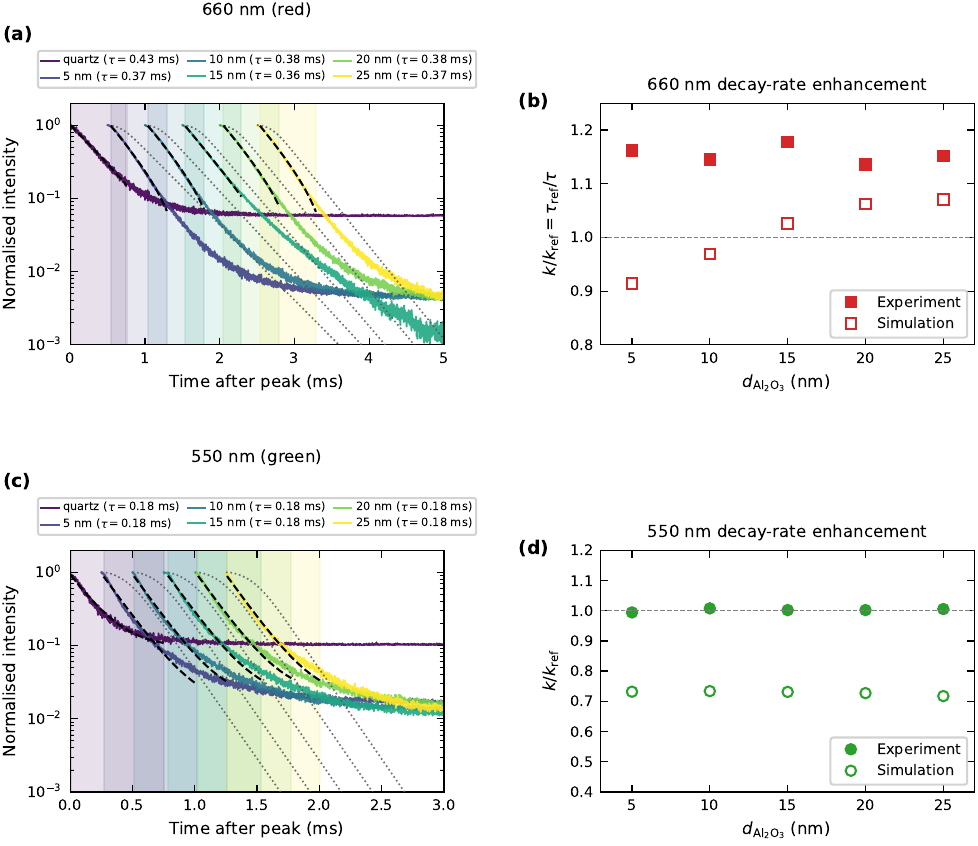}
  \caption{%
    \textbf{Time-resolved photoluminescence and decay-rate
    modulation.}
    (a,c)~Experimental normalised decay curves (shown from peak,
    rise dynamics excluded, translated horizontally for clarity)
    at \SI{660}{\nano\meter} (red band) and \SI{550}{\nano\meter}
    (green band) on quartz and for $d = 5$--\SI{25}{\nano\meter}.
    Black dashed: KWW fits (windows in SI
    Sec.~\hyperref[sec:SI_extended_decays]{S4.3}); dotted grey:
    coupled EM--kinetic simulation.
    (b,d)~Decay-rate enhancement
    $k/k_{\rm ref} = \tau_{\rm ref}/\tau$ vs.\ $d$: experiment
    (filled) and simulation (open).
    The red band shows a uniform $\approx +15\%$ enhancement,
    reproduced by the monotonic simulated trend (simulated
    $\approx 0.92, 0.97, 1.03, 1.07, 1.07$.
    The green lifetime is experimentally invariant
    ($|k/k_{\rm ref} - 1| < 1\%$) whereas the simulation gives
    $k/k_{\rm ref} \approx 0.73$; this offset is diagnosed by the
    systematic elimination tests of SI
    Sec.~\hyperref[sec:SI_extended_green]{S4.2}.
    See Sec.~\ref{sec:trpl}.
  }
  \label{fig:decays}
\end{figure}

Time-resolved PL measurements under pulsed \SI{980}{\nano\meter}
excitation (\SI{100}{\micro\second} pulses, \SI{10}{\hertz}
repetition rate) provide the most stringent test of the
electromagnetic--kinetic framework, because the decay rate is
a single-observable quantity directly proportional to the
LDOS\cite{NovotnyHecht} and therefore exposes any inaccuracy in
the predicted Purcell factors without the additional layers of
the rate-equation model.
Figure~\ref{fig:decays} reports the experimental decay traces
at \SI{660}{\nano\meter} (red band) and \SI{550}{\nano\meter}
(green band) for the five spacer thicknesses
$d = 5, 10, 15, 20, \SI{25}{\nano\meter}$ together with the
quartz reference, the corresponding effective lifetimes extracted
by Kohlrausch--Williams--Watts (KWW) fits with adaptive fit-window
selection (SI Sec.~\hyperref[sec:SI_extended_decays]{S4.3}; in
practice the stretching exponent $\beta \to 1$ for all samples,
and $\tau$ is the effective $1/e$ lifetime), and the decay-rate
enhancement
$k/k_{\rm ref} = \tau_{\rm ref}/\tau$ relative to the quartz
reference.
The simulated decay traces, obtained by numerically integrating
the rate equations with the FDTD-extracted
$F_{\rm rad}^{kj}$, $F_{\rm tot}^{kj}$ and $f_{\rm exc}$ as
inputs (SI Sec.~\hyperref[sec:SI_theory]{S2.2}), are overlaid in
Fig.~\ref{fig:decays}a, c as dotted grey curves; the simulated
$k/k_{\rm ref}$ values are reported as open symbols in
Fig.~\ref{fig:decays}b, d.

\paragraph{Red band: uniform enhancement reproduced.}
The experimental red-band decay-rate enhancement
$k/k_{\mathrm{ref}}=\tau_{\mathrm{ref}}/\tau$
(Fig.~\ref{fig:decays}b, filled squares) shows a uniform
shortening of the lifetime relative to the quartz reference
across the whole spacer range
($\tau \approx 0.36$--\SI{0.38}{\milli\second} vs.\
$\tau_{\rm ref} = \SI{0.43}{\milli\second}$), corresponding to
$k/k_{\mathrm{ref}} \approx 1.14$--$1.18$ with no significant
spacer dependence
($k/k_{\mathrm{ref}}\approx 1.16,\,1.15,\,1.18,\,1.14,\,1.15$ at
$d = 5, 10, 15, 20, \SI{25}{\nano\meter}$).

This experimental red-band enhancement ($\approx +15\%$) is consistent with the 
monotonic simulated trend ($\approx 10\%$ from
$k/k_{\rm ref}^{660} \approx 0.92$ at $d = \SI{5}{\nano\meter}$ to
$\approx 1.07$ at $d = \SI{25}{\nano\meter}$).
The framework thus captures the order of magnitude and the
global $d$-dependence of the LDOS modulation set by the broad
\SI{670}{\nano\meter} resonance.

\paragraph{Green band: model overshoots a perfectly invariant lifetime.}
In sharp contrast to the red band, the experimental green-band
lifetime is invariant within $\pm 1\%$ for all $d$:
$\tau_{\rm green} = \SI{0.18}{\milli\second}$ on every Au/Al$_2$O$_3$
sample and on the quartz reference (Fig.~\ref{fig:decays}c, d,
filled green circles).
This experimental band-selectivity is striking on its own
merits: the corrugated platform engineers the LDOS at the red
transition without perturbing the green transition,
confirming the spectral isolation of the \SI{670}{\nano\meter}
plasmonic resonance and providing a clean band-selective design
lever beyond the red channel alone.
Flat intermediate controls measured from the same UCNP stock
(bare SU8, flat Au, SU8+Au; SI Sec.~\hyperref[sec:SI_controls]{S4.6})
confirm that this is a property of the corrugated resonance and
the spacer: bare gold in direct contact with the UCNPs quenches
both bands non-selectively and \emph{raises} the green/red ratio
($+27\%$), i.e.\ in the opposite direction to the device, so mere
gold proximity cannot reproduce the band-selective red
engineering.

The coupled simulation, however, predicts a substantial
reduction of the green decay rate, with
$k/k_{\rm ref}^{\rm sim} \approx 0.73$ across the full $d$ range
(Fig.~\ref{fig:decays}d, open green circles), corresponding to
an over-prediction of the green-band $\tau$ by $\sim 35\%$.
This is the most visible disagreement of the framework with
experiment in the entire data set, and it is directly correlated
with the over-predicted extinction feature near
\SI{550}{\nano\meter} in the linear response
(Fig.~\ref{fig:platform}e) discussed in
Sec.~\ref{sec:platform}.

\paragraph{Ridge-tip smoothing and other diagnostic tests.}
The leading candidate explanation for the over-predicted
green-band feature is the idealised half-ellipse ridge profile
used in the FDTD model, which has sharper crests than the
smoother profile of the fabricated samples and would generate
an over-confined near-field feature at \SI{550}{\nano\meter}
that does not survive smoothing.
We tested this hypothesis directly by replacing the upper
$h_{\rm round}$ of each bump with a circular arc tangent to
the half-ellipse flanks with $C^1$ continuity, and repeating
the full LDOS sweep for
$h_{\rm round} \in \{0, 5, 10\}~\si{\nano\meter}$ at fixed
$d \in \{5, 10, 15, 20, 25\}~\si{\nano\meter}$ (the
$h_{\rm round}=0$ run is the baseline used throughout the main
text; see SI Sec.~\hyperref[sec:SI_extended_green]{S4.2} for
geometric definitions and full results).
Across the entire $d$ range and on both bands, the
surface-averaged radiative Purcell factor changes by only
$1$--$3\%$ between $h_{\rm round}=0$ and
$h_{\rm round}=\SI{10}{\nano\meter}$:
$\langle F_{\rm rad}^{660}\rangle$ stays within $1.04 \to 1.13$
at $d = \SI{5}{\nano\meter}$ and within $1.59 \to 1.62$ at
$d = \SI{25}{\nano\meter}$, while
$\langle F_{\rm rad}^{550}\rangle$ remains within
$1.02 \to 1.07$ at $d = \SI{5}{\nano\meter}$ and within
$1.10 \to 1.09$ at $d = \SI{25}{\nano\meter}$.
The associated decay-rate enhancement
$k/k_{\rm ref}^{550}$ is essentially unaffected
(Fig.~S2 of Sec.~\hyperref[sec:SI_extended_green]{S4.2}),
remaining within $0.71$--$0.75$ for all $h_{\rm round}$.
Apex shape on the scale of \SI{5}{\nano\meter}--\SI{10}{\nano\meter}
is therefore \emph{not} the dominant cause of the green-band
over-prediction: closing the residual $\sim 25\%$
$k/k_{\rm ref}^{550}$ gap requires corrections of much larger
amplitude than what apex smoothing alone can produce on this
geometry.

This negative result is informative, and we extended it to the
remaining accessible corrections (SI
Sec.~\hyperref[sec:SI_extended_green]{S4.2}). Making the gold
permittivity grain-boundary-aware (a Kreibig free-electron
size correction, $L = \SI{30}{\nano\meter}$) adds loss
predominantly in the \emph{red}
($\Delta\mathrm{Ext} \approx +0.02$ near
\SI{670}{\nano\meter} vs.\ $+0.005$ near \SI{550}{\nano\meter},
as expected from $\varepsilon'' \propto \lambda^3$) and changes
the green LDOS by only $\approx 3\%$ -- an order of magnitude
short of the $\sim 25\%$ correction required (Fig.~S3). 
Doubling the interband loss \emph{increases} the simulated
\SI{550}{\nano\meter} extinction -- the wrong direction, since
the model already over-predicts the feature(Fig.~S4).
Randomising the
ridge heights and widths and averaging over an ensemble of
realisations (inhomogeneous broadening) leaves the
\SI{550}{\nano\meter} feature unchanged to within a few percent (Fig.~S5).
 None of these
corrections closes the green residual.
Taken together, these tests establish that the simulated linear
extinction is the genuine, converged response of the idealised,
normal-incidence, bare-metal structure; its greater spectral
\emph{sharpness} relative to the measured extinction therefore
reflects measurement-versus-model factors -- the experiment
integrates over a finite-aperture cone of incidence angles that
smooths the angle-dispersive grating resonances a
normal-incidence simulation renders sharply -- rather than a
deficiency of the electromagnetic physics. For the decay
dynamics, the green manifold is predominantly non-radiative
($\approx 82\%$ from the calibrated rates), so its decay rate
is intrinsically far less sensitive to the LDOS than the
radiatively efficient red band, consistent with the measured
green invariance; the remaining gap is bounded by the
surface-averaged single-emitter LDOS approximation and by
green-side non-radiative/cross-relaxation channels not included
in the present six-level kinetic model.

\paragraph{Multi-observable validation and diagnostic role.}
The framework reproduces simultaneously
(i) the linear extinction resonance at \SI{670}{\nano\meter}
(Fig.~\ref{fig:platform}d, e),
(ii) the amplitude ($\pm 10\%$) of the red-band decay-rate
enhancement and the global increasing trend with $d$
(Fig.~\ref{fig:decays}b, monotonic in the simulation),
(iii) the qualitative trend and rough magnitude of the
power-density-dependent emission enhancement
(Fig.~\ref{fig:emission}c), and
(iv) the band-selective structure of the LDOS modification
(measurable amplitude at the red band, near-zero amplitude
at the green band).
The remaining quantitative discrepancies -- the
$\sim 1.5\times$
overestimation of the absolute green/red ratio, and the
$\sim 25\%$ over-prediction of the green-band rate reduction
-- are not separate failures but a single, internally
consistent over-prediction of short-wavelength spectral
features. The systematic elimination developed in this section
shows that neither apex shape, grain-boundary damping,
geometric disorder nor interband loss is a dominant lever on
this residual, re-orienting the search towards
measurement-versus-model factors (finite-aperture angular
averaging of the linear extinction) and additional green-side
kinetic channels not captured by the six-level model.
This is the diagnostic role of the framework: rather than
serving merely as a black-box predictor, it eliminates several
candidate hypotheses cleanly and ranks the remaining ones by
physical plausibility, providing a roadmap for further
development of both the geometric/material model and the
kinetic scheme.
For applications, the implication is that the framework is
quantitatively reliable for the amplitude and global structure
of LDOS-mediated effects when the underlying plasmonic feature
is broad and geometrically robust (here, the red Er$^{3+}$
band), and semi-quantitative for the precise $d$-dependent
fine structure of features sensitive to material details that
the canonical FDTD inputs do not yet describe.

\section{Conclusions}
\label{sec:conclusions}

We have studied a corrugated SU8/Au/Al$_2$O$_3$ grating coated
with a dense NaYF$_4$:Yb(20\%),Er(5\%) UCNP monolayer as a
band-selective platform for LDOS engineering of Yb/Er
upconversion, and have introduced a coupled
electromagnetic--kinetic framework to relate the simulated
electromagnetic environment to the multiple spectroscopic
observables of the platform.
The key findings, formulated as design rules for plasmonic
upconversion architectures, are summarised below.

\textbf{Band-selective LDOS engineering of the red Er$^{3+}$
transition.}
The platform supports a broad plasmonic LDOS resonance at the
red \SI{660}{\nano\meter} Er$^{3+}$ transition, producing a
decay-rate enhancement of $\approx 15\%$ as the Al$_2$O$_3$ spacer
thickness is varied between $d = 5$ and \SI{25}{\nano\meter},
while the green
$^2H_{11/2}/^4S_{3/2} \to {}^4I_{15/2}$ transition is left
experimentally invariant ($|k/k_{\rm ref} - 1| < 1\%$).
The pump field at \SI{980}{\nano\meter} is moderately and
monotonically suppressed below the free-space reference at all
spacer thicknesses, with no resonant enhancement: the platform
is a clean LDOS-only design lever, free of the pump-side
coupling that would otherwise have to be deconvolved from the
spectroscopic observables.
Tuning $d$ thus controls the green/red emission balance
\emph{independently} of the pump efficiency.

\textbf{Framework scope and predictive accuracy.}
The coupled framework reproduces, simultaneously: the linear
extinction resonance at \SI{670}{\nano\meter}; the amplitude
($\approx 10\%$) of the red-band decay-rate modulation and its
global monotonic trend with $d$; the order of magnitude and
dominant decreasing trend of the green/red intensity ratio;
and the qualitative magnitude of the
power-density-dependent emission enhancement.
The framework does \emph{not} reproduce two features of the
data:
the absolute magnitude of the green/red ratio (over-predicted
by $\sim 1.5\times$); and the experimentally invariant
green-band lifetime (the simulation predicts a $\sim 25\%$
reduction).
The framework is therefore quantitatively predictive for the
amplitude and global structure of LDOS-mediated effects when
the underlying plasmonic feature is broad and geometrically
robust (here, the red Er$^{3+}$ band), and semi-quantitative
for the precise $d$-dependent fine structure of features
sensitive to material details that the canonical FDTD inputs
(idealised half-ellipse geometry, Johnson--Christy gold
permittivity, six-level kinetic model with the parameters of
SI Table~S1) do not yet describe.

\textbf{Diagnostic by elimination.}
We tested the accessible candidate explanations for the
green-band over-prediction in turn -- ridge-tip smoothing,
grain-boundary (free-electron) damping of the gold, geometric
ridge disorder with inhomogeneous broadening, and enhanced
interband loss -- across the full $d$ range (SI
Sec.~\hyperref[sec:SI_extended_green]{S4.2}).
None closes the residual: apex smoothing and geometric disorder
move the surface-averaged radiative Purcell factor by only a
few percent (leaving $k/k_{\rm ref}^{550}$ within
$0.71$--$0.75$, far short of the $\sim 25\%$ correction needed);
grain-boundary damping acts predominantly on the red band and
shifts the green LDOS by only $\approx 3\%$; and interband loss
moves the \SI{550}{\nano\meter} feature in the wrong direction.
The simulated extinction is thus the genuine, converged
response of the idealised normal-incidence bare-metal
structure, and the residual is re-oriented to
measurement-versus-model factors (finite-aperture angular
averaging) and to additional green-side non-radiative channels
not captured by the six-level kinetic model.
This diagnosis by elimination is the most useful methodological
output of the framework: it cleanly rules out the accessible
electromagnetic and material corrections and orients the next
round of refinement towards the measurement/angular-averaging
treatment of the extinction and kinetic-level (green-side
channel) extensions of the model.

\textbf{Outlook.}
The methodology extends straightforwardly to more complex
architectures -- disordered metasurfaces, photonic crystal
slabs, gap-mode cavities -- and to other multilevel emitters
(Tm$^{3+}$/Yb$^{3+}$, Ho$^{3+}$/Yb$^{3+}$, Pr$^{3+}$ systems).
Combining the present band-selective LDOS platform with extreme
single-emitter Purcell regimes achievable in gap-mode plasmonic
nanocavities\cite{Wu2019,Chen2022NatPhoton} may enable
plasmonic-upconversion architectures with simultaneously
enhanced absorption, accelerated radiative decay and improved
out-coupling.
More immediately, closing the green-band sim/exp gap on the
present geometry -- by treating the finite-aperture angular
averaging of the measured extinction and by adding the missing
green-side kinetic channels to the rate equations -- would turn
the framework from a semi-quantitative
diagnostic into a fully predictive design tool for the next
generation of plasmonic-UCNP architectures.

\section{Methods}
\label{sec:methods}

\subsection{Sample Fabrication}
SU8 gratings were defined by UV interference lithography
(period \SI{400}{\nano\meter}) on quartz substrates.
Gold (\SI{50}{\nano\meter}) was deposited by thermal evaporation
at a base pressure below \SI{e-6}{\torr}.
Al$_2$O$_3$ spacers of 5, 10, 15, 20 and \SI{25}{\nano\meter}
were grown by ALD (trimethylaluminium/water, \SI{150}{\degreeCelsius})
at a rate of \SI{0.1}{\nano\meter} per cycle.
NaYF$_4$:Yb(20\%),Er(5\%) UCNPs (hexagonal phase,
$\sim$\SI{20}{\nano\meter} lateral size) were deposited by
spin-coating from hexane solution (\SI{3000}{rpm}, 60~s).\cite{Wang2010Nature,Wu2009PNAS}

\subsection{Optical Characterisation}
Transmission and reflection spectra were recorded with a
fibre-coupled spectrometer (Avantes AvaSpec-3648) and a
halogen white-light source in a normal-incidence configuration.
Steady-state PL spectra were acquired under \SI{980}{\nano\meter}
CW excitation (0.015--\SI{0.23}{\watt}, fibre-coupled diode laser)
focused to a $\sim\SI{50}{\micro\meter}$ spot, corresponding to
power densities of 300--\SI{9000}{\watt\per\centi\meter\squared}.
Time-resolved PL was measured using a pulsed \SI{980}{\nano\meter}
diode (\SI{100}{\micro\second} pulse, \SI{10}{\hertz} repetition rate)
and a photomultiplier tube with photon-counting electronics.
Decay curves were fitted to the Kohlrausch--Williams--Watts
stretched-exponential model with adaptive fit-window selection,
as detailed in
SI Sec.~\hyperref[sec:SI_extended_decays]{S4.3};
the stretching exponent was found to be $\beta \to 1$ for all
samples, so the reported $\tau$ corresponds to the effective
$1/e$ lifetime.

\subsection{FDTD Simulations}
All simulations were performed with Tidy3D
v2.10 (Flexcompute).\cite{Tidy3D}
The geometry consists of $N = 15$ half-ellipse-profile SU8 bumps
(period \SI{400}{\nano\meter}, semi-axes $b_{\rm SU8} = \SI{35}{\nano\meter}$
[amplitude] and $a_{1/2} = \SI{150}{\nano\meter}$ [half-width],
parametric form $(x, z) = (a_{1/2}\cos\theta,\, b_{\rm SU8}\sin\theta)$)
with a \SI{10}{\percent}
position jitter, conformally coated with \SI{50}{\nano\meter}
Au and $d$~nm Al$_2$O$_3$.
Perfectly matched layer (PML) boundary conditions are applied in $x$; absorbing
(Maxwellian absorber, 80 layers) boundary conditions are used
in $z$ to accommodate infinite flat layers; periodic conditions
are imposed in $y$ (2D geometry).
The Au dispersion is taken from Johnson and Christy tabulated
data;\cite{JohnsonChristy1972} SU8 and Al$_2$O$_3$ are treated
as non-dispersive with $n = 1.60$ and $n = 1.629$,
respectively.
A minimum of 20 grid points per wavelength is used globally,
with a 2~nm override mesh around the Au layer.
LDOS simulations are performed at the original 5-point spacer
sweep ($d = 5, 10, 15, 20, \SI{25}{\nano\meter}$); pump-field
simulations are extended by one point at $d = \SI{30}{\nano\meter}$
to confirm the asymptotic behaviour of
$\langle f_{\rm exc}\rangle$ at large $d$.
A parametric apex-smoothing diagnostic
($h_{\rm round} \in \{5, 10\}~\si{\nano\meter}$ in addition to
the baseline $h_{\rm round} = 0$) is performed across the same
five $d$ values; details of the smoothed-ridge geometry,
parametric sweep methodology and post-processing workflow are
given in
SI Sec.~\hyperref[sec:SI_FDTD]{S3} and
\hyperref[sec:SI_extended_green]{S4.2}.

\section*{Supporting Information}
The Supporting Information is here below.

Additional experimental details, materials, and methods, including
grating-substrate patterning, gold evaporation, Al$_2$O$_3$
atomic-layer deposition, and UCNP synthesis and monolayer deposition;
theoretical framework, including definitions of the electromagnetic
quantities and their FDTD extraction, the six-level
Yb$^{3+}$/Er$^{3+}$ rate-equation model with energy-level diagram,
and tabulated kinetic parameters; FDTD implementation details,
including geometry, boundary conditions, discretisation, monitor
configuration, and the smoothed-apex profile; extended simulation
results, including ridge-tip-smoothing, grain-boundary-damping,
interband-loss and ridge-disorder diagnostics,
Kohlrausch--Williams--Watts fits and complete decay curves for all
spacer thicknesses, power-dependent enhancement, pump-field
enhancement maps at \SI{980}{\nano\meter}, and flat-structure
controls (PDF)

\section*{Funding Sources}
Y.Z.\ and X.L.\ acknowledge support from Singapore Ministry of
Education (T2EP10224-0004) and RIE2025 Manufacturing, Trade and
Connectivity (MTC) Programmatic Fund (M24N7c0092).
M.G.C., S.R.\ and R.A.L.V.\ acknowledge support from the
French Agence Nationale de la Recherche (ANR).
M.G.C.\ and A.M.\ acknowledge support from the Spanish
MCIN/AEI/10.13039/501100011033 (PID2022-142025NB-I00).

\clearpage
\newpage
\begin{center}
{\LARGE\textbf{Supporting Information}}
\end{center}
\vspace{1em}

\let\addcontentsline\svaddcontentsline
\setcounter{secnumdepth}{-1}
\setcounter{tocdepth}{3}
\setcounter{figure}{0}
\renewcommand{\thefigure}{S\arabic{figure}}
\setcounter{table}{0}
\renewcommand{\thetable}{S\arabic{table}}
\setcounter{equation}{0}
\renewcommand{\theequation}{S\arabic{equation}}

\tableofcontents

\section{S1.\ Fabrication Details}
\label{sec:SI_fab}

\subsection{S1.1.\ Grating Substrate}
SU8 gratings were produced by UV laser interference lithography
using a \SI{325}{\nano\meter} He-Cd laser and a Lloyd's mirror
configuration, yielding a grating period of
$a = \SI{400 \pm 5}{\nano\meter}$.
After exposure, the SU8 layer was developed in PGMEA for 60~s
and hard-baked at \SI{150}{\degreeCelsius} for 30~min,
producing quasi-sinusoidal ridges on \SI{1}{\milli\meter}-thick
fused-silica substrates.
The ridge amplitude was measured by atomic force microscopy to be
\SI{35 \pm 3}{\nano\meter}.

\subsection{S1.2.\ Gold Deposition}
Au (\SI{50}{\nano\meter}) was deposited by resistive thermal
evaporation (Edwards Auto~306) at a base pressure of
$\leq \SI{5e-7}{\torr}$ and a deposition rate of
\SI{0.05}{\nano\meter\per\second}, ensuring conformal coverage
of the sinusoidal profile.
A \SI{2}{\nano\meter} titanium adhesion layer was first deposited.

\subsection{S1.3.\ Al$_2$O$_3$ Spacer by Atomic Layer Deposition (ALD)}
Al$_2$O$_3$ spacers were grown by atomic layer deposition (ALD) in a Beneq TFS~200 reactor
at \SI{150}{\degreeCelsius} using trimethylaluminium (TMA, pulse
\SI{0.2}{\second}, purge \SI{5}{\second}) and water vapour
(pulse \SI{0.5}{\second}, purge \SI{5}{\second}).
The growth rate was calibrated to
\SI{0.100}{\nano\meter} per cycle.
Target spacer thicknesses of 5, 10, 15, 20 and \SI{25}{\nano\meter}
were obtained with 50, 100, 150, 200 and 250 cycles, respectively.
Film thickness was verified on flat reference Si wafers by
spectroscopic ellipsometry.

\subsection{S1.4.\ UCNP Synthesis and Deposition}
NaYF$_4$:Yb(20\%),Er(5\%) UCNPs in the hexagonal ($\beta$) phase
were synthesised following a literature co-precipitation
method.\cite{UCNPsynthesis}
Transmission electron microscopy (TEM) characterisation showed hexagonal platelet morphology with a
lateral dimension of \SI{22 \pm 3}{\nano\meter} and a height of
\SI{15 \pm 2}{\nano\meter}.
UCNPs dispersed in hexane (\SI{5}{\milli\gram\per\milli\liter})
were spin-coated at \SI{3000}{rpm} for 60~s on the
ALD-coated gratings, yielding a dense, nearly close-packed
monolayer as confirmed by scanning electron microscopy (SEM) (Fig.~1c of the main text).

\section{S2.\ Theoretical Framework}
\label{sec:SI_theory}

\subsection{S2.1.\ EM Quantities: Definitions and FDTD Extraction}

The spontaneous emission rate of a quantum emitter with dipole
moment $\bm{\mu}$ at position $\mathbf{r}$ and angular frequency
$\omega$ in a linear dispersive environment is related to the
dyadic Green tensor $\mathbf{G}(\mathbf{r},\mathbf{r};\omega)$
by\cite{NovotnyHecht}
\begin{equation}
  \gamma(\mathbf{r},\omega)
  = \frac{2\omega^2}{\hbar\varepsilon_0 c^2}\,
    \bm{\mu}^\ast \cdot
    \operatorname{Im}\bigl\{\mathbf{G}(\mathbf{r},\mathbf{r};\omega)\bigr\}
    \cdot \bm{\mu}.
  \label{eq:gamma_G}
\end{equation}
In finite-difference time-domain (FDTD) practice, a classical point dipole source of polarisation
$\hat{\alpha}$ ($\alpha = x,y,z$) is placed at $\mathbf{r}$ and
the time-integrated power flowing through a closed surface
$\mathcal{S}$ enclosing the source is recorded as the total
emitted power $P_{\mathrm{tot}}^{(\alpha)}(\mathbf{r},\omega)$.
The total Purcell factor is then\cite{Esteban2008}
\begin{equation}
  F_{\mathrm{tot}}^{(\alpha)}(\mathbf{r},\omega)
  = \frac{P_{\mathrm{tot}}^{(\alpha)}(\mathbf{r},\omega)}
         {P_{\mathrm{tot},0}^{(\alpha)}(\mathbf{r},\omega)},
  \label{eq:Ftot_def}
\end{equation}
where $P_{\mathrm{tot},0}$ is the power emitted in the reference
environment (flat SU8 substrate without Au or Al$_2$O$_3$,
air above).
A second flux monitor of lateral size
$(L_x - \lambda_{\mathrm{max}}) \times \infty \times
 (L_z - \lambda_{\mathrm{max}})$ centred on the dipole
captures the radiated (far-field propagating) power
$P_{\mathrm{rad}}^{(\alpha)}$, defining the radiative Purcell
factor and the antenna efficiency\cite{Fischer2011OE,LiuLei2015}
\begin{align}
  F_{\mathrm{rad}}^{(\alpha)}(\mathbf{r},\omega)
    &= \frac{P_{\mathrm{rad}}^{(\alpha)}(\mathbf{r},\omega)}
            {P_{\mathrm{rad},0}^{(\alpha)}(\mathbf{r},\omega)},
  \label{eq:Frad_def}\\
  \eta_a(\mathbf{r},\omega)
    &= \frac{P_{\mathrm{rad}}(\mathbf{r},\omega)}
            {P_{\mathrm{tot}}(\mathbf{r},\omega)}.
  \label{eq:eta_def}
\end{align}

For randomly oriented emitters (isotropic UCNPs) in the
two-dimensional (2D) simulation plane ($xz$), the
orientation-averaged Purcell factors are approximated as
\begin{equation}
  \bar{F}_{\mathrm{tot}}(\mathbf{r},\omega)
  \approx \frac{2\,F_{\mathrm{tot}}^{(x)} + F_{\mathrm{tot}}^{(z)}}{3},
  \label{eq:Fiso}
\end{equation}
and similarly for $\bar{F}_{\mathrm{rad}}$, using the 2D mirror
symmetry $F^{(y)} \approx F^{(x)}$.
The surface-averaged Purcell factors entering the kinetic model
are obtained by averaging $\bar{F}$ over all sampled UCNP positions
along the Al$_2$O$_3$ surface profile.

The environment-induced nonradiative power is
\begin{equation}
  P_{\mathrm{env,nr}} = P_{\mathrm{tot}} - P_{\mathrm{rad}},
  \label{eq:Penvnr}
\end{equation}
and the corresponding environment-induced nonradiative rate
reads\cite{Fischer2011OE}
\begin{equation}
  \gamma_{\mathrm{env},kj}
  = (F_{\mathrm{tot},kj} - F_{\mathrm{rad},kj})\,
    \gamma^{0}_{\mathrm{rad},kj}.
  \label{eq:gamma_env}
\end{equation}

\subsection{S2.2.\ Rate-Equation Model}

\begin{figure}[H]
  \centering
  \includegraphics[width=0.7\textwidth]{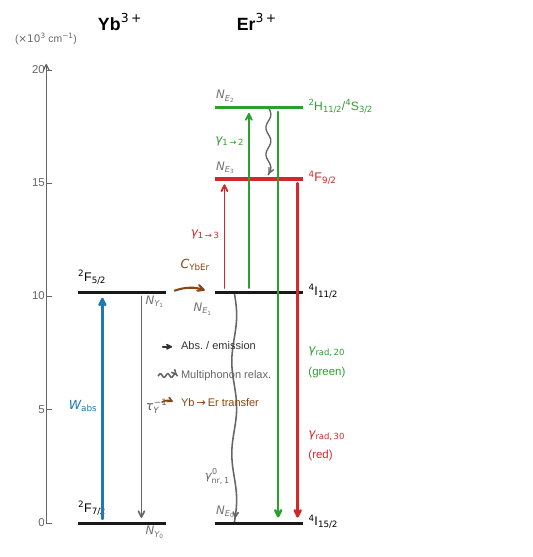}
  \caption{%
    Simplified six-level Yb$^{3+}$/Er$^{3+}$ energy-level diagram
    used in the rate-equation model.
    Solid upward arrows: radiative/absorption transitions.
    Wavy arrows: multiphonon relaxation.
    Curved arrows: Yb$\to$Er energy transfer ($C_{\mathrm{YbEr}}$).
    The model neglects Er$^{3+}$ levels above $E_2$ (green manifold).
  }
  \label{fig:SI_levels}
\end{figure}

The simplified six-level Yb/Er scheme retains the following
populations (see level diagram, Fig.~\ref{fig:SI_levels}):
$N_{Y_0}$ (Yb $^2$F$_{7/2}$, ground),
$N_{Y_1}$ (Yb $^2$F$_{5/2}$, excited),
$N_{E_0}$ (Er $^4$I$_{15/2}$, ground),
$N_{E_1}$ (Er $^4$I$_{11/2}$),
$N_{E_2}$ (Er $^2$H$_{11/2}$/$^4$S$_{3/2}$, green),
$N_{E_3}$ (Er $^4$F$_{9/2}$, red).
All populations are normalised to unity;
concentrations enter through effective rate coefficients
(see Sec.~\ref{sec:SI_theory}.S2.3 below).

Conservation gives
\begin{equation}
  N_{Y_0} + N_{Y_1} = 1, \qquad
  N_{E_0} + N_{E_1} + N_{E_2} + N_{E_3} = 1.
  \label{eq:conservation}
\end{equation}
The rate equations are
\begin{align}
  \frac{dN_{Y_1}}{dt}
  &= W_{\mathrm{abs}}\,N_{Y_0}
    - C_{\mathrm{YbEr}}\,N_{Y_1}\,N_{E_0}
    - \frac{N_{Y_1}}{\tau_{Y}},
  \label{eq:NY1}\\
  \frac{dN_{E_1}}{dt}
  &= C_{\mathrm{YbEr}}\,N_{Y_1}\,N_{E_0}
    - \bigl(\gamma_{\mathrm{rad},10}
            + \gamma^{0}_{\mathrm{nr},1}
            + \gamma_{1\to 2}
            + \gamma_{1\to 3}\bigr)\,N_{E_1},
  \label{eq:NE1}\\
  \frac{dN_{E_2}}{dt}
  &= \gamma_{1\to 2}\,N_{E_1}
    - \bigl(\gamma_{\mathrm{rad},20}
            + \gamma^{0}_{\mathrm{nr},2}
            + \gamma_{\mathrm{env},20}\bigr)\,N_{E_2},
  \label{eq:NE2}\\
  \frac{dN_{E_3}}{dt}
  &= \gamma_{1\to 3}\,N_{E_1}
    - \bigl(\gamma_{\mathrm{rad},30}
            + \gamma^{0}_{\mathrm{nr},3}
            + \gamma_{\mathrm{env},30}\bigr)\,N_{E_3},
  \label{eq:NE3}\\
  \frac{dN_{E_0}}{dt}
  &= -C_{\mathrm{YbEr}}\,N_{Y_1}\,N_{E_0}
    + \bigl(\gamma_{\mathrm{rad},10}
            + \gamma^{0}_{\mathrm{nr},1}\bigr)\,N_{E_1}
    + \bigl(\gamma_{\mathrm{rad},20}
            + \gamma^{0}_{\mathrm{nr},2}
            + \gamma_{\mathrm{env},20}\bigr)\,N_{E_2}
    \nonumber\\
  &\quad
    + \bigl(\gamma_{\mathrm{rad},30}
            + \gamma^{0}_{\mathrm{nr},3}
            + \gamma_{\mathrm{env},30}\bigr)\,N_{E_3}.
  \label{eq:NE0}
\end{align}
Here $W_{\mathrm{abs}} = f_{\mathrm{exc}}\,W_{\mathrm{abs}}^{(0)}\,I_{\mathrm{pump}}$
is the Yb ground-state absorption rate, enhanced by
$f_{\mathrm{exc}}$ relative to the reference.
The environment-modified radiative rates are
\begin{equation}
  \gamma_{\mathrm{rad},kj} = F_{\mathrm{rad},kj}\,\gamma^{0}_{\mathrm{rad},kj},
  \quad
  \gamma_{\mathrm{env},kj}
    = (F_{\mathrm{tot},kj} - F_{\mathrm{rad},kj})\,
      \gamma^{0}_{\mathrm{rad},kj}.
  \label{eq:gamma_env_rate}
\end{equation}

The instantaneous emitted intensities are
\begin{equation}
  I_{\mathrm{green}}(t) \propto \gamma_{\mathrm{rad},20}\,N_{E_2}(t),
  \qquad
  I_{\mathrm{red}}(t) \propto \gamma_{\mathrm{rad},30}\,N_{E_3}(t),
  \label{eq:intensities}
\end{equation}
and the green-over-red steady-state ratio is
\begin{equation}
  \frac{G}{R} = \frac{\gamma_{\mathrm{rad},20}\,N_{E_2}^{\mathrm{ss}}}
                     {\gamma_{\mathrm{rad},30}\,N_{E_3}^{\mathrm{ss}}}.
  \label{eq:GoverR}
\end{equation}

Transient decay curves are obtained by solving
Eqs.~\eqref{eq:NY1}--\eqref{eq:NE0} numerically (BDF method,
relative tolerance $10^{-4}$, absolute tolerance $10^{-12}$)
for $t > 0$ after setting $W_{\mathrm{abs}} = 0$ with the
steady-state populations as initial conditions.
The simulated decay traces shown in Fig.~4a, c of the main
text are taken from these solutions starting at the peak (rise
dynamics omitted for visualisation, consistently with the
experimental traces).

\subsection{S2.3.\ Parameter Values and Justification}

Table~\ref{tab:params} lists all kinetic parameters used in
the simulations.
The reference radiative lifetimes
$\tau^0_{\rm rad,green} = \tau^0_{\rm rad,red} = \SI{1}{\milli\second}$
are taken from the literature\cite{TorresVera2021,Fischer2011JAP},
and the intrinsic non-radiative rates are calibrated on the
measured lifetimes on bare quartz so that the unperturbed
$\tau$ is reproduced exactly.

\begin{table}[H]
	\centering
	\caption{%
		Kinetic parameters for
		$\beta$-NaYF$_4$:Yb(20\%),Er(5\%) UCNPs.
		All rate parameters correspond to free-space emission
		in a low-phonon-energy fluoride host or are calibrated
		on measured lifetimes on bare quartz.
		Population variables are normalised to unity; effective
		rate coefficients incorporate the concentration ratio
		$[{\rm Er}]/[{\rm Yb}] = 0.25$.
	}
	\label{tab:params}
	\footnotesize
	\setlength{\tabcolsep}{4pt}
	\renewcommand{\arraystretch}{1.15}
	\begin{tabularx}{\linewidth}{@{}l l X@{}}
		\toprule
		Symbol & Value & Source / justification \\
		\midrule
		\multicolumn{3}{@{}l}{\textit{Yb$^{3+}$ parameters}} \\
		$\tau_Y = 1/W_{Y_1}^0$
		& \SI{2.0}{ms}
		& Lit.\cite{TorresVera2021,Fischer2011JAP} \\
		$W_{\mathrm{abs}}^{(0)}$
		& $C_{\mathrm{YbEr}}\,([\mathrm{Er}]/[\mathrm{Yb}])$ s$^{-1}$
		& Calibrated so $I_{\mathrm{pump}}=1 \Leftrightarrow I_{\mathrm{sat}}$ \\
		$I_{\mathrm{sat}}$ (Yb)
		& \SI{3000}{W\,cm^{-2}}
		& Derived from $\sigma_Y = \SI{1.7e-20}{\centi\meter\squared}$
		and $\tau_Y$\cite{TorresVera2021} \\
		\midrule
		\multicolumn{3}{@{}l}{\textit{Energy transfer}} \\
		$K_2$ (Yb$\to$Er, $E_1$)
		& \SI{5e-17}{cm^3\,s^{-1}}
		& Ref.\cite{TorresVera2021} \\
		$C_{\mathrm{YbEr}} = K_2\,N_{\rm Yb}\times([\rm Er]/[\rm Yb])$
		& \SI{3.45e4}{s^{-1}}
		& Derived ($N_{\rm Yb} = \SI{2.76e21}{\centi\meter^{-3}}$) \\
		$K_4$ (Yb$\to$Er, $E_2$ via ETU)
		& $0.5\,K_2$ cm$^3$\,s$^{-1}$
		& Ref.\cite{TorresVera2021} \\
		$\gamma_{1\to 2}$
		& $1.73\times10^4$ s$^{-1}$
		& Derived \\
		$\gamma_{1\to 3}$
		& $W_{43} = 1.6\times10^4$ s$^{-1}$
		& Multi-phonon relaxation (see below) \\
		$C_{\mathrm{CR}}$
		& 0 s$^{-1}$
		& Absorbed into $\gamma^0_{\mathrm{nr},2}$ (see text) \\
		\midrule
		\multicolumn{3}{@{}l}{\textit{Er$^{3+}$ radiative rates (free space)}} \\
		$\gamma^0_{\mathrm{rad},20}$ (green, $E_2\to E_0$)
		& $10^3$ s$^{-1}$
		& Lit.\cite{TorresVera2021,Fischer2011JAP}; $\tau^0_{\rm rad,green}=1$~ms \\
		$\gamma^0_{\mathrm{rad},30}$ (red, $E_3\to E_0$)
		& $10^3$ s$^{-1}$
		& Lit.\cite{TorresVera2021}; $\tau^0_{\rm rad,red}=1$~ms \\
		$\gamma^0_{\mathrm{rad},10}$ ($E_1\to E_0$, \SI{980}{\nano\meter})
		& $500$ s$^{-1}$
		& Lit.\cite{Fischer2011JAP} \\
		\midrule
		\multicolumn{3}{@{}l}{\textit{Er$^{3+}$ intrinsic non-radiative rates}} \\
		$W_{43}$ (multi-phonon $E_2\to E_3$)
		& \SI{1.6e4}{s^{-1}}
		& From $\eta_0 = W_{40}/(W_{40}+W_{43})=0.06$\cite{TorresVera2021} \\
		$\gamma^0_{\mathrm{nr},1} = W_{21}$
		& $W_{43} = 1.6\times10^4$ s$^{-1}$
		& $W_{32}=W_{21}=W_{43}$\cite{TorresVera2021} \\
		$\gamma^0_{\mathrm{nr},2}$ (green)
		& $4556$ s$^{-1}$
		& Calibrated on measured $\tau_{\rm green}=\SI{180}{\micro\second}$
		on quartz \\
		$\gamma^0_{\mathrm{nr},3}$ (red)
		& $1326$ s$^{-1}$
		& Calibrated on measured $\tau_{\rm red}=\SI{430}{\micro\second}$
		on quartz \\
		\midrule
		\multicolumn{3}{@{}l}{\textit{Concentrations}} \\
		$N_{\rm Yb}$
		& \SI{2.76e21}{cm^{-3}}
		& $u_V=\SI{0.1086}{\nano\meter\cubed}$ (hexagonal NaYF$_4$)\cite{Mackenzie} \\
		$N_{\rm Er}$
		& \SI{6.90e20}{cm^{-3}}
		& Same unit cell \\
		$[{\rm Er}]/[{\rm Yb}]$
		& 0.25
		& By construction \\
		\bottomrule
	\end{tabularx}
\end{table}

\textbf{Calibration of non-radiative rates.}
The total decay rates of the green and red emitting manifolds
in the absence of a nanostructure (UCNPs on quartz) are
measured directly:
$\tau_{\rm green}^{\rm exp} = \SI{180}{\micro\second}$ and
$\tau_{\rm red}^{\rm exp} = \SI{430}{\micro\second}$.
Since the total decay rate is
$\gamma_{\rm tot} = \gamma^0_{\rm rad} + \gamma^0_{\rm nr}$,
the intrinsic non-radiative rates are obtained by subtraction:
\begin{equation}
  \gamma^0_{\rm nr,2}
    = \frac{1}{\tau_{\rm green}^{\rm exp}}
      - \gamma^0_{\rm rad,20}
    = \SI{4556}{\per\second},
  \label{eq:gnr2}
\end{equation}
\begin{equation}
  \gamma^0_{\rm nr,3}
    = \frac{1}{\tau_{\rm red}^{\rm exp}}
      - \gamma^0_{\rm rad,30}
    = \SI{1326}{\per\second}.
  \label{eq:gnr3}
\end{equation}
These values include all intrinsic decay channels (surface
quenching, multi-phonon relaxation, cross-relaxation) and are
treated as effective, sample-specific parameters.
The cross-relaxation coefficient $C_{\rm CR}$ is set to zero
because its effect is absorbed into $\gamma^0_{\rm nr,2}$ in
this single-sample calibration.

\textbf{Pump saturation.}
The Yb$^{3+}$ saturation intensity is estimated from the
absorption cross-section $\sigma_Y = \SI{1.7e-20}{\centi\meter\squared}$
and the excited-state lifetime $\tau_Y$:
\begin{equation}
  I_{\rm sat}
    = \frac{h\nu}{\sigma_Y\,\tau_Y}
    \approx \SI{3}{\kilo\watt\per\centi\meter\squared},
\end{equation}
consistent with the sub-quadratic power dependence ($n < 2$)
observed experimentally over the range
300--\SI{9000}{\watt\per\centi\meter\squared}.
The effective pump absorption coefficient is calibrated so that
$I_{\rm pump} = 1$ in the model corresponds to
$I = I_{\rm sat}$, giving
$W_{\rm abs}^{(0)} = C_{\rm YbEr}\times[\rm Er]/[\rm Yb]
= \SI{3.45e4}{\per\second}$.

\section{S3.\ FDTD Implementation}
\label{sec:SI_FDTD}

\subsection{S3.1.\ Software and Version}
All FDTD simulations were performed using Tidy3D v2.10
(Flexcompute, Inc.).\cite{Tidy3D}
Tidy3D is a cloud-based, GPU-accelerated FDTD solver with a
Python API. Jobs were submitted and managed via the
\texttt{tidy3d.web} batch interface.

\subsection{S3.2.\ Geometry}

\subsubsection{Structure}
The simulation geometry consists of $N = 15$ half-ellipse-profile
SU8 bumps centred symmetrically around $x = 0$ with period
$a = \SI{400}{\nano\meter}$, semi-axes $b_{\rm SU8} = \SI{35}{\nano\meter}$
(amplitude) and $a_{1/2} = \SI{150}{\nano\meter}$ (half-width).
The bump profile is generated parametrically as
$(x, z) = (a_{1/2}\cos\theta,\, b_{\rm SU8}\sin\theta)$ with
$\theta \in [\pi, 0]$, equivalently
$z(x) = b_{\rm SU8}\sqrt{1 - (x/a_{1/2})^2}$ for $|x|\le a_{1/2}$.
The bumps sit on an infinite SU8 base layer of thickness
$h_{\rm SU8} = \SI{80}{\nano\meter}$.
A position jitter of $\pm 10\%$ of the period (random seed 42) is
applied to reproduce the fabrication disorder observed in SEM images.
Au shells of thickness $b_{\rm Au} = \SI{50}{\nano\meter}$ conformally
coat the bumps.
An Al$_2$O$_3$ layer of thickness $d$ conformally coats the Au.

The bump profiles are implemented as \texttt{PolySlab} objects
in Tidy3D, with vertices computed from the half-ellipse profile
plus uniform vertical offset for the Au and Al$_2$O$_3$ shells.
Shapely's \texttt{buffer(0)} healing is applied to all polygons
before constructing the \texttt{PolySlab} objects to avoid
degenerate vertices.

\subsubsection{Reference Geometry}
The reference geometry (used to normalise emitted power and
pump field) consists of a flat SU8 substrate ($h_{\rm SU8} = \SI{80}{\nano\meter}$)
with air above.
This choice corresponds to the experimental reference sample
(UCNPs on bare quartz/SU8 without Au or Al$_2$O$_3$) and
ensures that the computed Purcell factors directly measure the
modification of emission rates by the metallic nanostructure.

\subsubsection{UCNP Sampling Positions}
UCNP positions are sampled along the Al$_2$O$_3$ outer surface
profile of the central bump ($x \approx 0$) with an arc-length
spacing of \SI{20}{\nano\meter}, placing each dipole at a distance
$r_{\rm UCNP} = \SI{15}{\nano\meter}$ above the surface along the
local outward normal.
By mirror symmetry ($x \to -x$), only the right half-period
($x \geq 0$) is sampled, giving 11 positions per simulation.
The same sampling grid is used at all spacer thicknesses to
ensure that the spatial resolution of the surface average is
independent of $d$.

\subsection{S3.3.\ Boundary Conditions}

\begin{itemize}
  \item $x$: perfectly matched layers (PML, 24 layers) absorb
    outgoing radiation from the finite-length structure.
  \item $y$: periodic boundary conditions (2D geometry,
    $L_y = 0$).
  \item $z$: Maxwellian absorbers (80 layers) are used rather
    than PML because the substrate and flat Au/Al$_2$O$_3$
    layers are infinite in $x$ and $y$ and cross the $z$
    boundaries; absorbers avoid the numerical instabilities
    that arise when PML boundaries intersect structures with
    non-zero tangential fields.\cite{TafloveHagness}
\end{itemize}

\subsection{S3.4.\ Spatial and Temporal Discretisation}

The FDTD grid uses Tidy3D's automatic grid generator
(\texttt{GridSpec.auto}) with a minimum of 20 grid points per
wavelength at the central simulation wavelength.
A \SI{2}{\nano\meter} mesh override is enforced throughout the
Au and Al$_2$O$_3$ layers to resolve the conformal shells.
For the emission simulations (broadband \SI{520} - \SI{700}{\nano\meter}),
the grid is set at the central frequency \SI{600}{\nano\meter};
for the pump simulation (\SI{980}{\nano\meter}), the grid is
sized accordingly.

The simulation run time is set to \SI{1}{\pico\second} with a
field shutoff criterion of $10^{-6}$ (relative energy decay).
For all converged simulations the field had decayed below this
threshold before the end of the run time, as verified by the
\texttt{final\_decay\_value} attribute.

\subsection{S3.5.\ Source and Monitor Configuration}

\subsubsection{Emission Simulations (LDOS Map)}
For each UCNP position and polarisation ($x$ or $z$), a
\texttt{PointDipole} source with a broadband Gaussian pulse
(central frequency $f_0 = c/\SI{610}{\nano\meter}$,
relative bandwidth $\Delta f/f_0 = 0.295$, covering
\SI{520}-\SI{700}{\nano\meter}) is placed at the UCNP
position.
Two \texttt{FluxMonitor} objects record the emitted powers:
\begin{itemize}
  \item \textit{P\textsubscript{tot} box}: size
    $4r_{\rm UCNP} \times \infty \times 4r_{\rm UCNP} =
    \SI{60}{\nano\meter} \times \infty \times \SI{60}{\nano\meter}$,
    centred on the dipole; captures total emitted power including
    ohmic losses.
  \item \textit{P\textsubscript{rad} box}: size
    $(L_x - \lambda_{\rm max}) \times \infty \times
     (L_z - \lambda_{\rm max})$, centred on the dipole in $z$;
    captures far-field radiated power.
\end{itemize}
The same source and monitor configuration is used in the
structured and reference geometries.
Purcell factors and antenna efficiencies are computed at 61
uniformly spaced frequencies from \SI{520}{\nano\meter} to
\SI{700}{\nano\meter}, and the values at \SI{550}{\nano\meter}
and \SI{660}{\nano\meter} are extracted by interpolation.

\subsubsection{Pump Simulations (\SI{980}{\nano\meter})}
A \texttt{PlaneWave} source is placed at
$z = z_{\rm max} + 0.6\lambda_{\rm pump}$ above the structure,
propagating downward ($-z$ direction) with TM polarisation
(E along $x$).
A \texttt{FieldMonitor} in the $xz$ plane covering the UCNP
zone records $E_x$ and $E_z$ at the pump frequency.
The pump enhancement $f_{\rm exc}(\mathbf{r}) =
|E(\mathbf{r})|^2/|E_{\rm ref}(\mathbf{r})|^2$
is evaluated at each UCNP position by nearest-neighbour
interpolation on the monitor grid.

\subsection{S3.6.\ Computational Cost}

Each LDOS simulation (one dipole polarisation, broadband)
was launched on the Tidy3D cloud.
For 11 positions $\times$ 2 polarisations, the structure batch
(22 jobs) plus reference batch (2 jobs), per d value, were run sequentially. 
The pump simulation (1 structure + 1 reference), per $d$ value,
were subsequently launched. 

\subsection{S3.7.\ Smoothed-Apex Geometry for the Ridge-Tip
            Diagnostic}
\label{sec:SI_FDTD_smoothed}

To test ridge-tip smoothing as a candidate correction for the
green-band over-prediction reported in Sec.~2.1 of
the main text and analysed in Sec.~\hyperref[sec:SI_extended_green]{S4.2}
below, we replace the upper $h_{\rm round}$ of each SU8 bump
by a circular arc tangent to the half-ellipse flanks with
$C^1$ continuity:
\begin{equation}
z(x) = b_{\rm SU8}\sqrt{1-(x/a_{1/2})^2} \;\to\;
       z_{\rm smooth}(x; h_{\rm round})
\quad \mathrm{for}\quad
z > z_{\rm match} = b_{\rm SU8} - h_{\rm round},
\label{eq:smoothing}
\end{equation}
with the matching point $(x_m, z_m)$ on the ellipse flank set by
$z_m = b_{\rm SU8} - h_{\rm round}$ and
$x_m = a_{1/2}\sqrt{1-(z_m/b_{\rm SU8})^2}$.
The arc radius $R_{\rm tip} = x_m / \sin\theta_m$ is a derived
quantity, where
$\theta_m = \arctan[(b_{\rm SU8}/a_{1/2})^2(x_m/z_m)]$ is the
tangent angle of the half-ellipse at the match point.
For the present geometry ($a_{1/2} = \SI{150}{\nano\meter}$,
$b_{\rm SU8} = \SI{35}{\nano\meter}$),
$h_{\rm round} = \SI{5}{\nano\meter}$ gives
$R_{\rm tip} \approx \SI{556}{\nano\meter}$ and
$h_{\rm round} = \SI{10}{\nano\meter}$ gives
$R_{\rm tip} \approx \SI{471}{\nano\meter}$.
The Au and Al$_2$O$_3$ shells are conformally regenerated on
the smoothed SU8 profile with the same thicknesses as the
nominal geometry; the surface-sampling positions for the
UCNP dipoles are recomputed on the new Al$_2$O$_3$ outer
profile (11 positions per simulation, same arc-length spacing
of \SI{20}{\nano\meter} as the baseline).
The ridge-tip-smoothing diagnostic sweep covers
$h_{\rm round} \in \{0, 5, 10\}~\si{\nano\meter}$ at
$d \in \{5, 10, 15, 20, \SI{25}{\nano\meter}\}$, for a total of
fifteen LDOS simulations sharing identical cloud settings.
The pump-field simulations are not repeated for
$h_{\rm round} > 0$: the pump-side response is essentially
homogeneous laterally on this platform (Sec.~2.2
of the main text), so apex shape on the scale of
\SI{5}{\nano\meter}--\SI{10}{\nano\meter} is not expected to
shift $\langle f_{\rm exc}\rangle$ significantly, and we hold
the baseline pump runs frozen across the diagnostic to isolate
the LDOS effect.

\section{S4.\ Extended Simulation Results and Diagnostic Analysis}
\label{sec:SI_extended}

\subsection{S4.1.\ Convergence and Antenna-Efficiency Trend}
\label{sec:SI_extended_convergence}

All structure and reference simulations reached the field
shutoff criterion of $10^{-6}$ (relative energy decay) well
before the end of the \SI{1}{\pico\second} run time, confirming
adequate convergence.
Doubling the spatial mesh resolution (from 20 to 40 grid points
per wavelength) changed the surface-averaged
$\langle F_{\rm rad}\rangle$ by less than 3\% at both wavelengths.
Halving the dipole sampling spacing (from \SI{20}{\nano\meter}
to \SI{10}{\nano\meter} along the Al$_2$O$_3$ surface) changed
the surface average by less than 2\%, confirming that 11
positions are sufficient to converge the surface integral.

\paragraph{Antenna-efficiency vs.\ spacer trend.}
The decrease of $\eta_a$ with $d$ for both bands
(Fig.~2b, main text) requires comment, because in simple
single-emitter--single-mirror geometries, $\eta_a$ typically
\emph{increases} with the emitter--metal distance
($\eta_a \to 1$ as $d \to \infty$) due to vanishing ohmic
loss\cite{NovotnyHecht}.
The opposite trend observed here reflects the corrugated
geometry of the platform: at small $d$, the radiative coupling
is dominated by a strongly hybridised plasmonic mode with
substantial dipole moment in the air half-space, channelling
emission into propagating plane waves; as $d$ increases, the
hybridised mode redshifts and weakens, so the relative weight
of dissipative pathways (lossy surface modes, bound modes
above the substrate) grows.
Within the spacer range investigated (\SI{5}{\nano\meter}
$\le d \le \SI{25}{\nano\meter}$), the metal is never effectively
"far away" -- all emitters are within the near-field of the
plasmonic mode -- and the $1/d^n$ scaling of single-emitter
quenching does not apply.
The numerical values of $\eta_a$ are nevertheless within the
range expected for emitters at metallic interfaces with thin
dielectric spacers\cite{Anger2006,Fischer2011OE}, and the
decrease is moderate (less than 20\% relative change at the
red band).

\subsection{S4.2.\ Ridge-Tip-Smoothing Diagnostic and the
            Green-Band Discrepancy}
\label{sec:SI_extended_green}

The most visible disagreement between the coupled framework
and experiment occurs at the green band.
At the green $^2H_{11/2}/^4S_{3/2} \to {}^4I_{15/2}$ transition,
the framework predicts
$k/k_{\rm ref}^{\rm sim} \approx 0.73$ across the full $d$ range
(Fig.~4d, main text, open circles), while the experimental
green-band lifetime is invariant at
$\tau_{\rm green}^{\rm exp} = \SI{0.18}{\milli\second}$ for all
$d$ ($|k/k_{\rm ref}^{\rm exp} - 1| < 1\%$).
This translates to a $\sim 25\%$ over-prediction of the
green-band $\tau$ that propagates into the steady-state
observables (the $\sim 1.5\times$ over-prediction of the $G/R$
ratio in Fig.~3b, main text).
At the red ${}^4F_{9/2} \to {}^4I_{15/2}$ transition, the
framework captures the amplitude of the modulation
($\pm 10\%$) and its global \emph{monotonic} increasing trend
with $d$
($k/k_{\rm ref}^{\rm sim} \approx 0.92, 0.97, 1.03, 1.07, 1.07$
at $d = 5, 10, 15, 20, \SI{25}{\nano\meter}$), in agreement with
the measured experimental data (uniform $\approx +15\%$.
The same idealised FDTD model also predicts an extinction
feature near \SI{550}{\nano\meter} in the linear response that
is suppressed by a factor of $\sim 2$ in the experimental
spectra (Fig.~1d vs.\ 1e, main text).
We test below, in turn, every accessible electromagnetic and
material correction to the idealised model -- ridge-tip
smoothing, grain-boundary (free-electron) damping, geometric
disorder with inhomogeneous broadening, and enhanced interband
loss -- and conclude with a clean negative result that
re-orients the diagnostic search.

\paragraph{Candidate physical effects not captured by the model.}
Four independent physical effects are not captured by the
idealised FDTD model:
\begin{enumerate}
  \item \textbf{Ridge-tip smoothing.}
    The half-ellipse profile used in the FDTD model has sharp
    apices that produce strongly confined near-fields and a
    correspondingly bright extinction feature at
    \SI{550}{\nano\meter}.
    The fabricated ridges, after development and hard-bake,
    have rounded tips: the upper $\sim 5$--\SI{10}{\nano\meter}
    of each ridge effectively follow a smoother profile that
    might suppress high-spatial-frequency near-field components
    and damp the short-wavelength resonance.
  \item \textbf{Grain-boundary damping in evaporated Au.}
    The Johnson--Christy permittivity\cite{JohnsonChristy1972}
    used in the FDTD model is fitted to single-crystalline bulk
    gold and underestimates the optical losses of thermally
    evaporated thin films, where grain boundaries
    (\SI{20}--\SI{50}{\nano\meter} grain size in our deposition)
    add scattering at the metal--air interface.
    Such damping is sometimes expected to affect short-wavelength
    plasmonic features; we test this expectation explicitly
    below.
  \item \textbf{Inhomogeneous broadening from ridge-position
    jitter.}
    Although our model already includes a $\pm 10\%$ position
    jitter of each ridge (Sec.~S3.2), the idealised flat-bump
    profile within each ridge does not account for amplitude
    fluctuations or width variations that would further smear
    sharp spectral features.
  \item \textbf{Enhanced interband loss.}
    Surface roughness and grain boundaries in evaporated gold
    can broaden the bound-electron (interband) transitions that
    dominate $\varepsilon''$ in the 520--\SI{560}{\nano\meter}
    region, an effect distinct from the free-electron
    (Drude/grain) correction above and acting precisely at the
    green band.
\end{enumerate}

\paragraph{Parametric diagnostic: ridge-tip smoothing sweep.}
We tested the first of these three corrections directly by
recomputing $\langle F_{\rm rad}^{550}\rangle$ and
$\langle F_{\rm rad}^{660}\rangle$ on smoothed-apex ridge
profiles defined by Eq.~\eqref{eq:smoothing} of
Sec.~\ref{sec:SI_FDTD_smoothed}, with
$h_{\rm round} \in \{0, 5, 10\}~\si{\nano\meter}$ at
$d \in \{5, 10, 15, 20, \SI{25}{\nano\meter}\}$
(15 LDOS simulations).
Two predictions of the framework are tested simultaneously:
\textit{(a)}~the green-band radiative Purcell factor should
drop from its idealised value at $h_{\rm round}=0$ towards
$\langle F_{\rm rad}^{550}\rangle \approx 1.0$ as
$h_{\rm round}$ increases, recovering the experimentally
observed green-band lifetime invariance; \textit{(b)}~the
red-band radiative Purcell factor should remain essentially
unaffected, given the broad and geometrically robust character
of the \SI{670}{\nano\meter} resonance.

\begin{figure}[H]
  \centering
  \includegraphics[width=\textwidth]{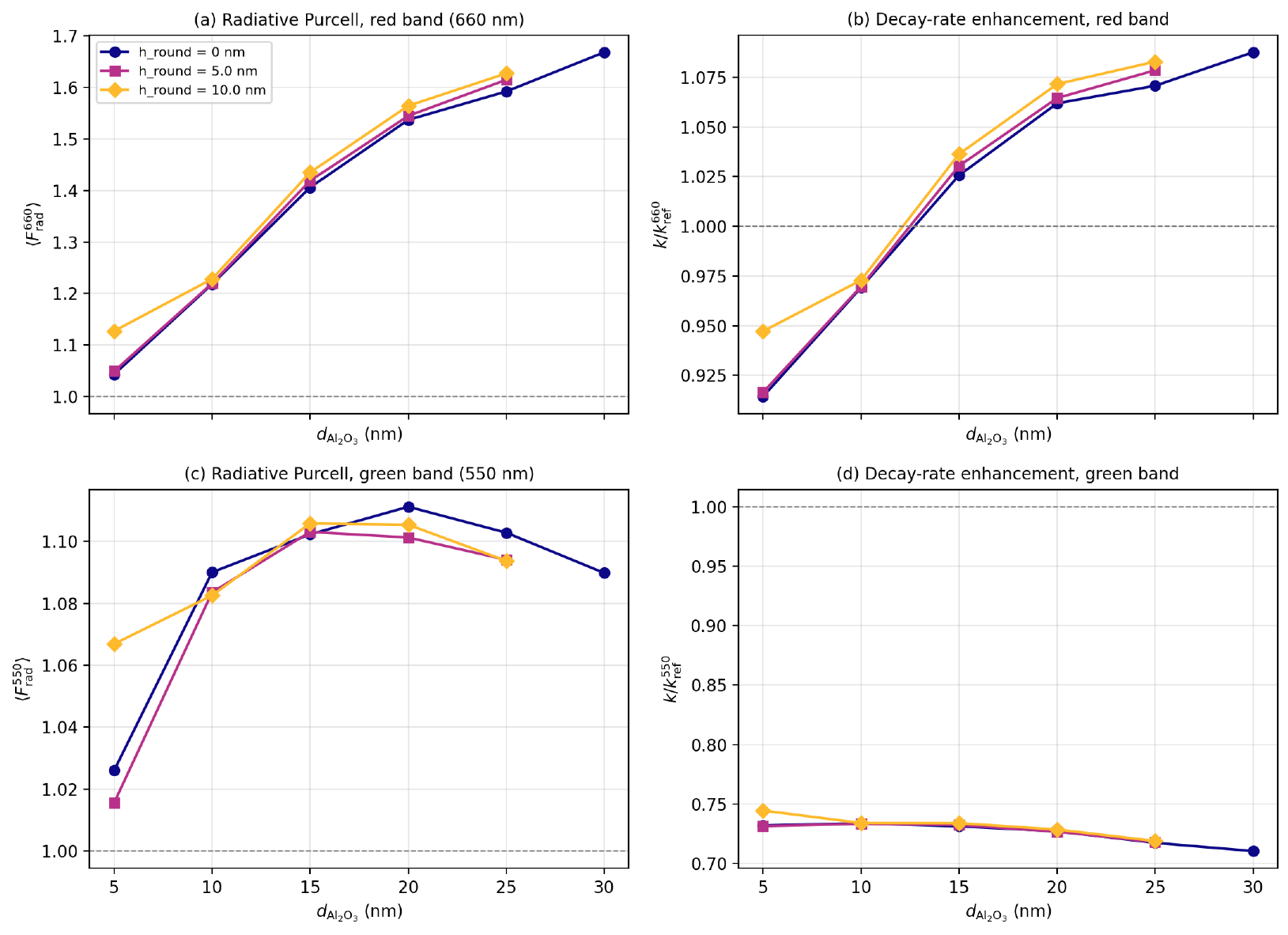}
  \caption{%
    \textbf{Ridge-tip-smoothing diagnostic.}
    Effect of replacing the upper $h_{\rm round}$ of each SU8
    bump by a $C^1$-tangent circular arc
    (Eq.~\eqref{eq:smoothing}) on the surface-averaged radiative
    Purcell factor and on the simulated decay-rate enhancement.
    (a)~$\langle F_{\rm rad}^{660}\rangle$ vs.\ $d$ for
    $h_{\rm round} \in \{0, 5, 10\}~\si{\nano\meter}$ (the
    rightmost point at $d = \SI{30}{\nano\meter}$ is shown for
    $h_{\rm round}=0$ only).
    (b)~$k/k_{\rm ref}^{660}$ vs.\ $d$ for the same three
    $h_{\rm round}$ values, computed from the time-resolved
    decays of the rate-equation model with the smoothed-apex
    Purcell factors as inputs.
    (c)~$\langle F_{\rm rad}^{550}\rangle$ vs.\ $d$.
    (d)~$k/k_{\rm ref}^{550}$ vs.\ $d$.
    Apex smoothing changes the surface-averaged radiative
    Purcell factor by only $1$--$3\%$ on either band; the
    decay-rate enhancement at the green band remains within
    $0.71$--$0.75$ for all $h_{\rm round}$, far short of the
    $\sim 25\%$ correction needed to match the experimentally
    invariant $k/k_{\rm ref}^{550} \approx 1$ (Fig.~4d, main
    text).
    Apex smoothing on the scale set by the lithography is
    therefore not the dominant correction.
  }
  \label{fig:SI_rtip}
\end{figure}

\paragraph{Result and interpretation.}
The diagnostic sweep is summarised in
Fig.~\ref{fig:SI_rtip}.
Across the entire $d$ range and on both bands, the
surface-averaged radiative Purcell factor changes by only
$1$--$3\%$ between $h_{\rm round}=0$ and
$h_{\rm round}=\SI{10}{\nano\meter}$.
At the red band (panel a), the three curves
$h_{\rm round} \in \{0, 5, 10\}~\si{\nano\meter}$ are
visually superimposed within the marker size from
$d = \SI{10}{\nano\meter}$ onwards; the only sub-percent
variation is at $d = \SI{5}{\nano\meter}$, where
$\langle F_{\rm rad}^{660}\rangle$ rises slightly with
$h_{\rm round}$ ($1.04 \to 1.13$), reflecting the modest
recession of the smoothed apex away from the metal.
At the green band (panel c), the three curves are also
indistinguishable to within $\sim 3\%$, with the maximum
near $d = \SI{15}{\nano\meter}$ retained at all
$h_{\rm round}$.
The decay-rate enhancements (panels b, d), computed by feeding
the smoothed-apex Purcell factors into the rate-equation model
of Sec.~\ref{sec:SI_theory}, follow accordingly:
$k/k_{\rm ref}^{660}$ remains within $\pm 1\%$ of the
$h_{\rm round}=0$ baseline at every $d$ (panel b), and
$k/k_{\rm ref}^{550}$ remains pinned at $0.71$--$0.75$ for all
$h_{\rm round}$ (panel d).

This is well short of what would be required to close the green-band residual: the
experimentally observed $k/k_{\mathrm{ref}}^{550} = 1.00 \pm 0.01$ requires a
$\sim$25\% upward correction of the simulated $0.73$, whereas apex smoothing on the
scale of 5\,nm--10\,nm delivers a $\leq 3\%$ effect. At the red band, apex smoothing
leaves the simulated $k/k_{\mathrm{ref}}^{660}(d)$ essentially unchanged -- the curve
remains a slightly translated version of the $h_{\mathrm{round}} = 0$ baseline --
consistent with the broad, geometrically robust character of the 670\,nm resonance and
with the uniform measured red-band enhancement. We conclude that ridge-tip smoothing on
the scale set by the lithography is not the dominant correction for the green-band
discrepancy on this platform.

\begin{figure}[H]
  \centering
  \includegraphics[width=0.8\textwidth]{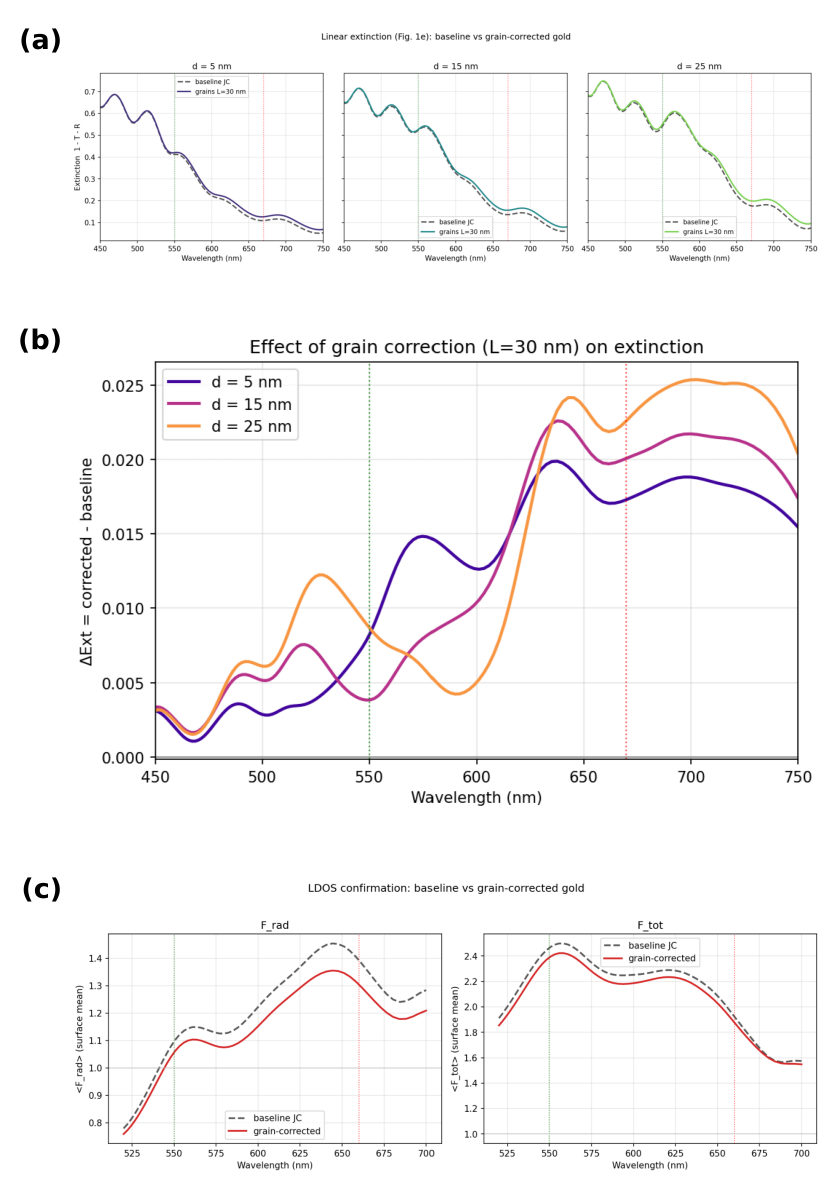}
  \caption{%
    \textbf{Grain-boundary (Drude) correction.}
    (a)~Linear extinction (baseline Johnson--Christy vs.\
    grain-corrected, $L = \SI{30}{\nano\meter}$) at
    $d = 5, 15, \SI{25}{\nano\meter}$; (b)~$\Delta\mathrm{Ext}$
    (corrected $-$ baseline), showing the red-dominant character
    of the added loss; (c)~confirmatory LDOS ($\langle F_{\rm
    rad}\rangle$, $\langle F_{\rm tot}\rangle$) at
    $d = \SI{15}{\nano\meter}$, with the green band changing by
    only $\approx 3$--$4\%$. The free-electron grain correction
    does not resolve the green residual.
  }
  \label{fig:SI_grain}
\end{figure}

\begin{figure}[H]
  \centering
  \includegraphics[width=0.8\textwidth]{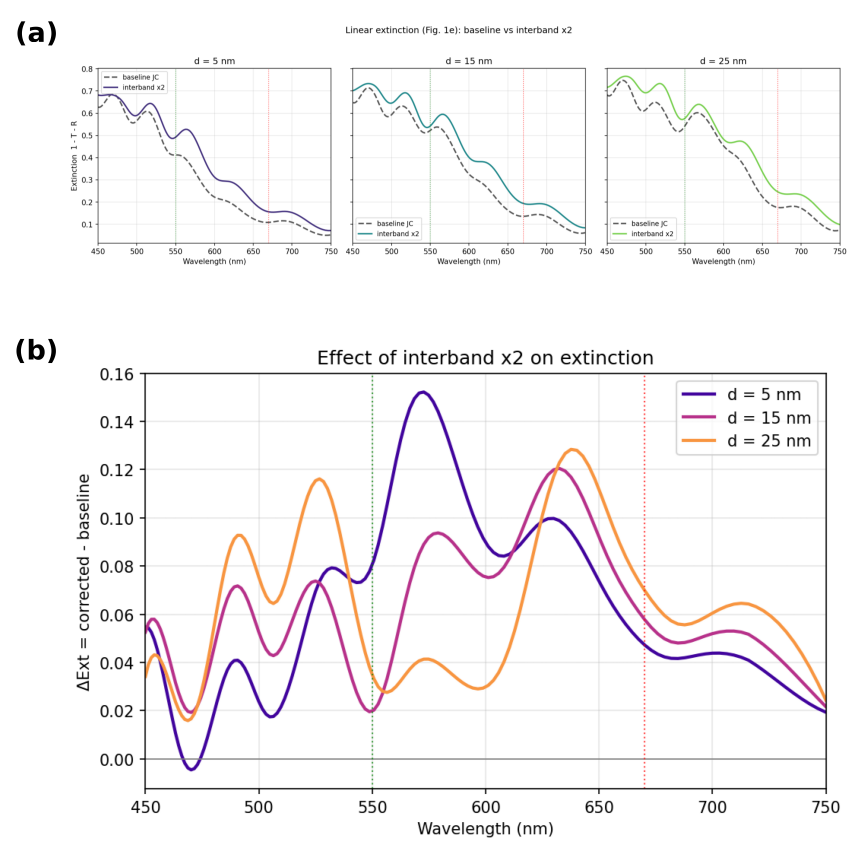}
  \caption{%
    \textbf{Enhanced interband loss.}
    (a)~Linear extinction (baseline vs.\ interband
    $\times 2$) at $d = 5, 15, \SI{25}{\nano\meter}$;
    (b)~$\Delta\mathrm{Ext}$ (corrected $-$ baseline). Doubling
    the interband loss \emph{increases} the \SI{550}{\nano\meter}
    extinction -- the opposite of what is required, since the
    idealised model already over-predicts the green feature.
  }
  \label{fig:SI_interband}
\end{figure}

\begin{figure}[H]
  \centering
  \includegraphics[width=0.8\textwidth]{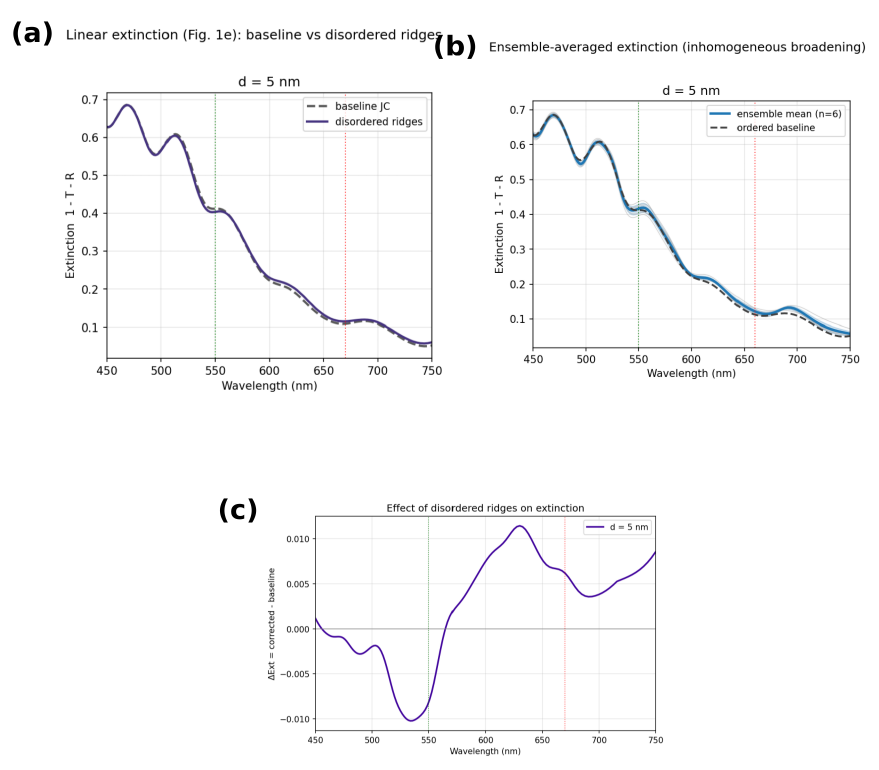}
  \caption{%
    \textbf{Geometric disorder and inhomogeneous broadening.}
    (a)~Extinction with disordered ridge heights/widths
    ($\pm 30\%$) vs.\ the ordered baseline;
    (b)~ensemble-averaged extinction over six realisations vs.\
    the ordered baseline; (c)~$\Delta\mathrm{Ext}$ for the
    disordered case. The \SI{550}{\nano\meter} feature is
    unchanged to within a few percent.
  }
  \label{fig:SI_disorder}
\end{figure}

\paragraph{Extended diagnostic: permittivity and disorder.}
We applied the same elimination logic to the three remaining
candidates. Making the gold permittivity grain-boundary-aware
(a Kreibig free-electron size correction with grain size
$L = \SI{30}{\nano\meter}$) adds extinction predominantly in
the \emph{red} ($\Delta\mathrm{Ext} \approx +0.017$--$0.025$
near \SI{670}{\nano\meter} vs.\ only $+0.004$--$0.009$ near
\SI{550}{\nano\meter}, Fig.~\ref{fig:SI_grain}), as expected
from the free-electron scaling $\varepsilon'' \propto \lambda^3$,
and a confirmatory LDOS run at $d = \SI{15}{\nano\meter}$ shows
the green radiative Purcell factor changing by only
$\approx 3.6\%$ -- an order of magnitude short of the
$\sim 25\%$ correction required. Randomising the ridge heights
and widths ($\pm 30\%$) and averaging over an ensemble of six
realisations leaves the \SI{550}{\nano\meter} extinction
unchanged to within a few percent
(Fig.~\ref{fig:SI_disorder}), so geometric disorder and
inhomogeneous broadening do not suppress the green feature.
Finally, doubling the interband loss (a causal Lorentz
oscillator near the 2.4~eV edge) \emph{increases} the simulated
\SI{550}{\nano\meter} extinction by up to $\Delta\mathrm{Ext}
\approx 0.08$--$0.15$ (Fig.~\ref{fig:SI_interband}) -- the
wrong direction, since the model already over-predicts the
feature.

\paragraph{Diagnostic implication.}
None of the four accessible corrections closes the green-band
residual: ridge-tip smoothing and geometric disorder move the
surface-averaged radiative Purcell factor by only a few
percent, grain-boundary (free-electron) damping acts
predominantly on the red band, and enhanced interband loss
moves the green feature in the wrong direction.
These negative results establish that the simulated linear
extinction is the genuine, converged response of the idealised,
normal-incidence, bare-metal structure: it reproduces the
resonance positions and the band structure correctly, and its
greater spectral \emph{sharpness} relative to the measured
extinction reflects a measurement-versus-model difference --
the experiment integrates over a finite-numerical-aperture cone
of incidence angles that smooths the angle-dispersive grating
resonances rendered sharply by a normal-incidence calculation
-- rather than a deficiency of the electromagnetic physics.
For the decay dynamics, the green manifold is predominantly
non-radiative ($\approx 82\%$ from the calibrated rates), so
its decay rate is intrinsically far less sensitive to the LDOS
than the radiatively efficient red band, consistent with the
measured green invariance; the residual
$k/k_{\rm ref}^{550}$ gap is therefore bounded by the
surface-averaged single-emitter LDOS approximation and by
green-side non-radiative/cross-relaxation channels not
described by the present six-level kinetic model. The remaining
direction for future refinement is thus the
measurement/angular-averaging treatment of the linear
extinction together with an extended green-side kinetic scheme,
rather than any of the bare-structure electromagnetic
corrections tested here.

\subsection{S4.3.\ Fitting Methodology and Complete Decay Curves}
\label{sec:SI_extended_decays}

\paragraph{Kohlrausch--Williams--Watts model with adaptive fit window.}
Experimental decay traces were fitted with the
Kohlrausch--Williams--Watts (KWW) stretched-exponential
form,\cite{Kohlrausch1854,Williams1970}
\begin{equation}
  I(t) = A\,\exp\!\left[-\bigl(t/\tau\bigr)^{\beta}\right] + C,
  \label{eq:KWW}
\end{equation}
which reduces to a mono-exponential decay when $\beta = 1$ and
generalises to non-exponential dynamics expected from
distributed environments otherwise.
The fit window is selected adaptively for each trace: the start
is set at the time after the peak when $I/I_{\rm max}$ first
drops below 0.95, and the end is set at the time when
$I/I_{\rm max}$ falls below $1.1\times$ the average noise level
in the long-time tail (9--\SI{10}{\milli\second}).
This procedure ensures that the fit excludes both the
ETU-mediated rise and the noise-dominated late tail.
Free parameters $(A,\tau,\beta,C)$ are obtained by non-linear
least-squares minimisation with bounds
$0 \le \beta \le 1$.
For all 12 traces (6 thicknesses $\times$ 2 bands), the
fitted $\beta \to 1$ within numerical tolerance, indicating
that the underlying decay is mono-exponential to the precision
of the measurement; $\tau$ is therefore the effective $1/e$
lifetime.

\paragraph{Fitted parameters.}
Table~\ref{tab:SI_decays} reports the fitted $\tau$, $\beta$,
fit window and resulting $k/k_{\rm ref}$ for each sample.
The reference $\tau_{\rm ref}$ is the value fitted on the
quartz sample under identical conditions, which makes
$k/k_{\rm ref}$ insensitive to systematic measurement biases
(e.g.\ minor pulse-shape variations or detection
non-linearity).

\begin{table}[H]
  \centering
  \caption{%
    Fitted decay parameters for all spacer thicknesses at the
    red and green bands.
    Fit windows refer to the time after the peak.
    The KWW $\beta$ exponent is reported to three decimal places
    to confirm $\beta \to 1$.
  }
  \label{tab:SI_decays}
  \begin{tabular}{lcccc}
    \toprule
    Sample & $\tau$ (ms) & $\beta$ & Fit window (ms) &
      $k/k_{\rm ref} = \tau_{\rm ref}/\tau$ \\
    \midrule
    \multicolumn{5}{l}{\textit{Red band, \SI{660}{\nano\meter}}} \\
    Quartz (ref) & 0.430 & 1.000 & 0.80--1.55 & 1.000 \\
    $d = \SI{5}{\nano\meter}$  & 0.370 & 1.000 & 0.85--1.60 & 1.162 \\
    $d = \SI{10}{\nano\meter}$ & 0.375 & 1.000 & 0.85--1.60 & 1.146 \\
    $d = \SI{15}{\nano\meter}$ & 0.365 & 1.000 & 0.85--1.60 & 1.178 \\
    $d = \SI{20}{\nano\meter}$ & 0.378 & 1.000 & 0.85--1.60 & 1.137 \\
    $d = \SI{25}{\nano\meter}$ & 0.373 & 1.000 & 0.85--1.60 & 1.152 \\
    \midrule
    \multicolumn{5}{l}{\textit{Green band, \SI{550}{\nano\meter}}} \\
    Quartz (ref) & 0.183 & 1.000 & 0.80--1.55 & 1.000 \\
    $d = \SI{5}{\nano\meter}$  & 0.184 & 1.000 & 0.81--1.56 & 0.994 \\
    $d = \SI{10}{\nano\meter}$ & 0.182 & 1.000 & 0.81--1.56 & 1.008 \\
    $d = \SI{15}{\nano\meter}$ & 0.183 & 1.000 & 0.81--1.56 & 1.002 \\
    $d = \SI{20}{\nano\meter}$ & 0.183 & 1.000 & 0.81--1.56 & 1.002 \\
    $d = \SI{25}{\nano\meter}$ & 0.182 & 1.000 & 0.81--1.56 & 1.005 \\
    \bottomrule
  \end{tabular}
\end{table}

\begin{figure}[H]
  \centering
  \includegraphics[width=\textwidth]{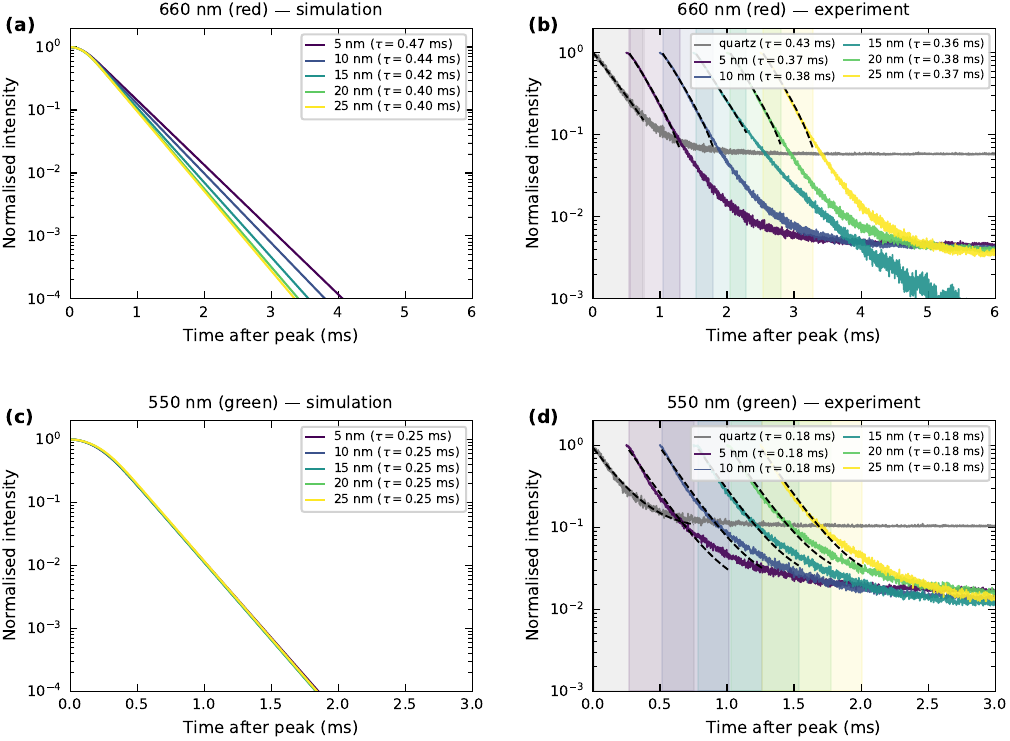}
  \caption{%
    Complete normalised decay curves for all spacer thicknesses
    $d = 5$, 10, 15, 20, \SI{25}{\nano\meter}.
    (a)~Simulated red band (\SI{660}{\nano\meter}).
    (b)~Experimental red band, with KWW fits as dashed lines
    and shaded fit windows.
    (c)~Simulated green band (\SI{550}{\nano\meter}).
    (d)~Experimental green band.
    Colourmap: viridis (purple = \SI{5}{\nano\meter},
    yellow = \SI{25}{\nano\meter}).
    Each trace is plotted from its peak (rise dynamics excluded);
    a horizontal time-shift has been applied for visual
    separation.
  }
  \label{fig:SI_decays_all}
\end{figure}

\paragraph{Complete decay curves.}
Figure~\ref{fig:SI_decays_all} shows the simulated and
experimental normalised decay curves at \SI{550}{\nano\meter}
(green band) and \SI{660}{\nano\meter} (red band) for all
$d$ values, with each trace normalised to its own peak and
plotted on a logarithmic vertical axis.
The red-band traces show a comparable shortening relative to
quartz across all spacers; the green-band traces are essentially
super-imposable.

\subsection{S4.4.\ Power-Dependent Enhancement and Reference Baseline}
\label{sec:SI_extended_power}

\begin{figure}[H]
  \centering
  \includegraphics[width=\textwidth]{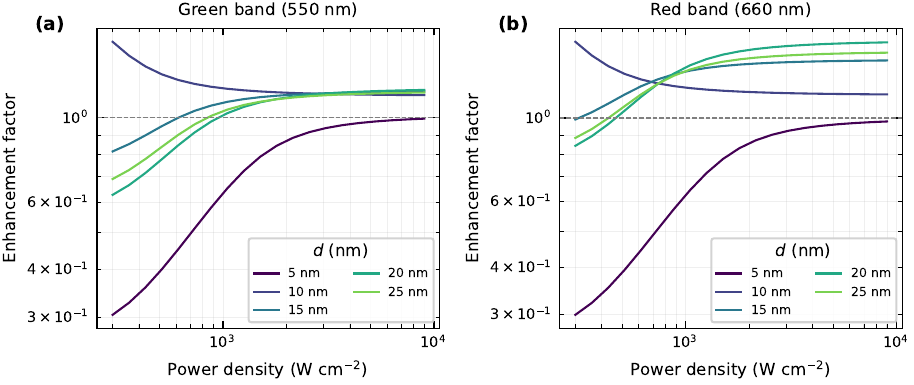}
  \caption{%
    Simulated emission enhancement (structure / flat-SU8
    reference) vs.\ pump-power density for
    $d = 5$--\SI{25}{\nano\meter}.
    (a)~Green band (\SI{550}{\nano\meter}).
    (b)~Red band (\SI{660}{\nano\meter}).
    Enhancement decreases with increasing irradiance in the
    Yb saturation regime
    ($I > I_{\rm sat} \approx \SI{3}{\kilo\watt\per\centi\meter\squared}$).
  }
  \label{fig:SI_enhancement}
\end{figure}

Figure~\ref{fig:SI_enhancement} shows the simulated emission
enhancement relative to the flat-SU8 reference as a function
of pump-power density for all $d$ values.
The enhancement decreases with increasing pump density in the
saturation regime, consistent with the experimental observation
of sub-quadratic power dependence (Fig.~3a, inset).

\paragraph{Reference-baseline correction.}
The experimentally measured enhancement factors (10--35$\times$
relative to UCNPs on bare quartz) are larger by roughly an
order of magnitude than the values predicted in
Fig.~\ref{fig:SI_enhancement} relative to the flat-SU8
reference.
The discrepancy is fully accounted for by two effects not
captured in our flat-SU8 reference:
\textit{(i)}~the experimental reference (UCNPs on bare quartz)
sits in a lower-index environment than UCNPs on flat SU8
($n_{\rm quartz} = 1.46$ vs.\ $n_{\rm SU8} = 1.60$), reducing
the local field at the emitter and thus lowering the LDOS
baseline by a factor estimated at 4--5 from a coarse flat-stack
calculation; and
\textit{(ii)}~the corrugated stack acts as an effective
anti-reflection coating at \SI{980}{\nano\meter}, increasing
the pump transmission into the UCNP layer by a further factor
of $\sim 2$ relative to flat SU8.
The product of these two factors brings the simulated and
experimental enhancement values into rough agreement, while
the relative ordering of the $d$ values and the qualitative
power dependence are faithfully reproduced.

\subsection{S4.5.\ Pump-Field Enhancement Maps for All Spacer Thicknesses}
\label{sec:SI_extended_maps}

\begin{figure}[H]
  \centering
  \includegraphics[width=\textwidth]{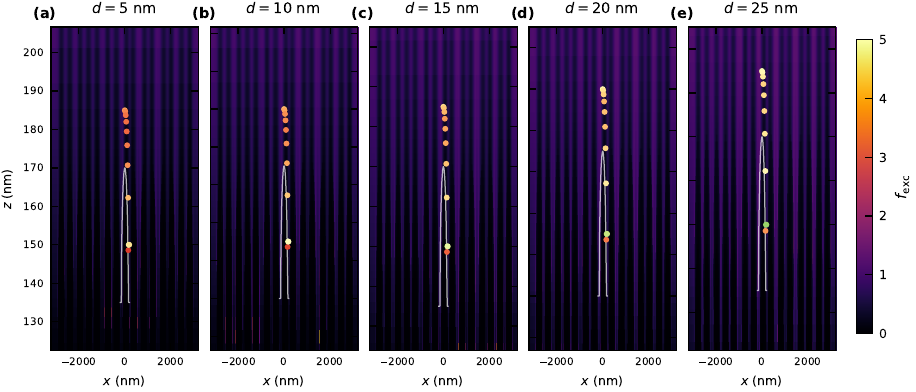}
  \caption{%
    Pump-field enhancement maps $f_{\rm exc}(x,z)$ at
    \SI{980}{\nano\meter} on a logarithmic colour scale
    for (a)~$d = \SI{5}{\nano\meter}$,
    (b)~$d = \SI{10}{\nano\meter}$,
    (c)~$d = \SI{15}{\nano\meter}$,
    (d)~$d = \SI{20}{\nano\meter}$,
    (e)~$d = \SI{25}{\nano\meter}$.
    The maps are spatially homogeneous to within weak vertical
    striations imprinted by the $\sim$\SI{400}{\nano\meter}-period
    grating; $f_{\rm exc}$ at the UCNP positions remains below or
    near unity at all $d$, with no localised hotspots.
    Coloured circles: UCNP sampling positions, colour-coded
    by local $f_{\rm exc}$ on the same logarithmic scale.
    White curves: Al$_2$O$_3$ surface profile.
  }
  \label{fig:SI_fexc_all}
\end{figure}

Figure~\ref{fig:SI_fexc_all} shows the pump-field enhancement
$f_{\rm exc}(x,z)$ maps at \SI{980}{\nano\meter} for the
five LDOS-sweep thicknesses
($d = 5, 10, 15, 20, \SI{25}{\nano\meter}$), on a logarithmic
colour scale.
At all spacer thicknesses, $f_{\rm exc}$ is suppressed below
unity over essentially the entire UCNP sampling region above
the bump apex; the field maps are spatially homogeneous to
within the weak vertical striations imprinted by the
$\sim$\SI{400}{\nano\meter}-period grating.
The trend in the surface-averaged
$\langle f_{\rm exc}\rangle$ reported in
Fig.~2c of the main text -- a monotonic
increase from $\approx 0.27$ at $d = \SI{5}{\nano\meter}$ to
$\approx 0.48$ at $d = \SI{25}{\nano\meter}$ -- reflects the
progressive recession of the UCNP layer from the metal as
$d$ grows, with no resonant pump-side enhancement on this
geometry.

\subsection{S4.6.\ Flat-Structure Controls: Contact Quenching
            vs.\ Band-Selective Resonance}
\label{sec:SI_controls}

\begin{figure}[H]
  \centering
  \includegraphics[width=\textwidth]{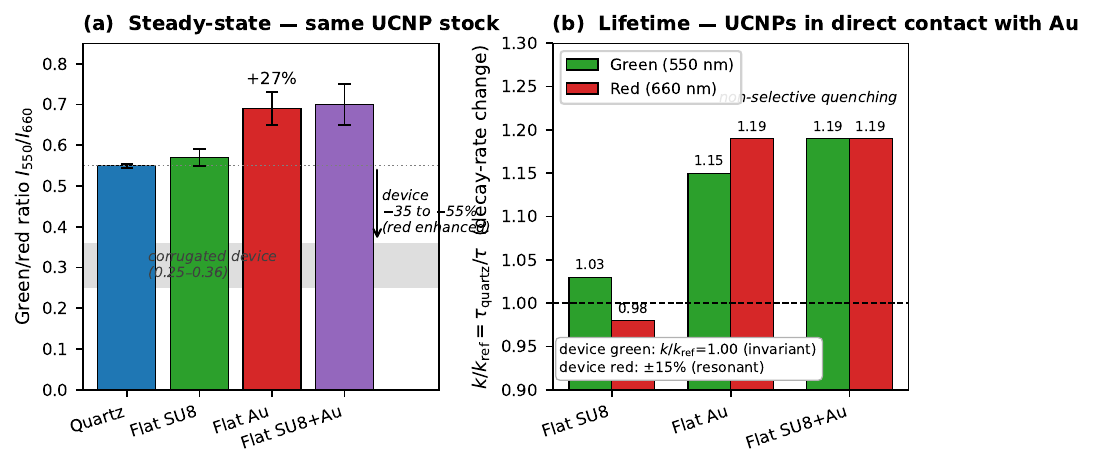}
  \caption{%
    \textbf{Flat-structure controls.}
    All data use the same UCNP stock as the device samples and
    the quartz reference of the main text.
    (a)~Steady-state green/red intensity ratio $I_{550}/I_{660}$
    (background-subtracted, \SI{9}{\kilo\watt\per\centi\meter\squared})
    for quartz, flat SU8, flat Au and flat SU8+Au (bars, mean
    $\pm$ s.d.\ over three positions); the shaded band marks the
    corrugated device range (0.25--0.36, Fig.~3b). Bare gold
    raises the ratio ($+27\%$), whereas the device lowers it:
    opposite signs for the same nanoparticles.
    (b)~Decay-rate change $k/k_{\rm ref} = \tau_{\rm quartz}/\tau$
    at 550~nm (green) and 660~nm (red) for the flat controls,
    UCNPs in direct contact with the metal. Bare Au shortens
    both bands by a comparable amount (non-selective contact
    quenching).
  }
  \label{fig:SI_controls}
\end{figure}

To isolate the contributions of the substrate, the metal and
the corrugated resonance, we characterised three flat
intermediate structures by steady-state and time-resolved
photoluminescence: bare SU8 on quartz, a \SI{50}{\nano\meter}
flat Au film, and a flat SU8+Au stack. These controls were
prepared from the \emph{same UCNP stock solution} used for the
device samples and are referenced to the \emph{same quartz
measurement} reported in the main text
($\tau_{\rm green} = \SI{180}{\micro\second}$,
$\tau_{\rm red} = \SI{430}{\micro\second}$;
$I_{550}/I_{660} = 0.55$). Steady-state spectra were recorded
at \SI{9}{\kilo\watt\per\centi\meter\squared} (the comparison
power of Fig.~3a). On the flat Au and SU8+Au controls the
UCNPs are in \emph{direct contact} with the gold (no
Al$_2$O$_3$ spacer), i.e.\ the maximal contact-quenching
configuration. Results are
summarised in Table~\ref{tab:SI_controls} and
Fig.~\ref{fig:SI_controls}.

\textbf{Substrate.} Flat SU8 reproduces the quartz reference
within scatter ($k/k_{\rm ref} = 1.03 / 0.98$ at $550/660$~nm;
$\Delta G/R = +5\%$), confirming that the SU8 index does not by
itself modify the LDOS appreciably.

\textbf{Bare gold: non-selective contact quenching.} With the
UCNPs directly on Au, both bands shorten by a comparable amount
($k/k_{\rm ref} = 1.15$ at 550~nm, $1.19$ at 660~nm). The
near-equal magnitude on the two bands identifies this as
broadband near-field quenching, carrying no spectral
selectivity; it is independent of the underlayer
(SU8+Au $\approx$ Au).

\textbf{Sign reversal of the green/red ratio.} The steady-state
ratio moves in \emph{opposite directions} for the flat metal
control and for the device: bare Au \emph{raises}
$I_{550}/I_{660}$ by $+27\%$ (red quenched more), whereas the
corrugated device \emph{lowers} it to 0.25--0.36 (Fig.~3b).
Since the same nanoparticles are used throughout, the device's
red enhancement cannot be attributed to the presence of gold:
it requires the grating plasmonic resonance spectrally aligned
with the red transition, together with the dielectric spacer.

\textbf{Role of the spacer at the green band.} Bare Au quenches
the green band ($k/k_{\rm ref} = 1.15$), yet the corrugated
device shows a perfectly invariant green lifetime
($|k/k_{\rm ref}-1| < 1\%$ for all $d$). The reconciliation is
spatial: the $\ge \SI{5}{\nano\meter}$ Al$_2$O$_3$ spacer and
the UCNP stand-off keep the green emitters beyond the
contact-quenching range, while the green transition is in any
case spectrally isolated from the \SI{670}{\nano\meter}
resonance. Decay parameters are single mono-exponential fits
per band ($R^2 > 0.99$); steady-state ratios are means over
three positions per sample.

\begin{table}[H]
  \centering
  \caption{%
    Flat-structure controls (same UCNP stock; quartz reference
    from the main text).
  }
  \label{tab:SI_controls}
  \begin{tabular}{lccccc}
    \toprule
    Configuration & $\tau_{\rm green}$ (\si{\micro\second}) &
      $k/k_{\rm ref}^{550}$ & $\tau_{\rm red}$ (\si{\micro\second}) &
      $k/k_{\rm ref}^{660}$ & $I_{550}/I_{660}$ \\
    \midrule
    Quartz (reference)        & 180 & 1.00 & 430 & 1.00 & 0.55 \\
    Flat SU8                  & 174 & 1.03 & 440 & 0.98 & 0.57 \\
    Flat Au (UCNPs on Au)     & 156 & 1.15 & 362 & 1.19 & 0.69 \\
    Flat SU8+Au (UCNPs on Au) & 151 & 1.19 & 362 & 1.19 & 0.70 \\
    Corrugated device ($d = 5$--\SI{25}{\nano\meter})
      & 180 & 1.00 & 365--378 & 1.14--1.18 & 0.25--0.36 \\
    \bottomrule
  \end{tabular}
\end{table}

\FloatBarrier

\FloatBarrier
\bibliographystyle{unsrt}
\bibliography{references}

\end{document}